\documentclass[10pt, oneside]{article}   	
\pdfoutput=1
\usepackage{geometry}                		
\usepackage{mathptmx}       
\usepackage{helvet}         
\usepackage{courier}        
\usepackage{type1cm}        
\usepackage{makeidx}         
\usepackage{graphicx}        
\usepackage{multicol}        
\usepackage[bottom]{footmisc}

\usepackage{array}
\usepackage{subfigure}
\usepackage{latexsym}
\usepackage{verbatim}
\usepackage{epsfig}
\usepackage{amsmath}
\usepackage{epstopdf}
\usepackage{times}
\usepackage{amsfonts}
\usepackage{graphics}
\usepackage{amssymb}
\usepackage{url}

\begin{document}

\title{Mining and Analyzing the Italian Parliament: \\ Party Structure and Evolution}

\author{Alessia Amelio \footnote{National Research Council of Italy (CNR), Institute for High Performance Computing and Networking (ICAR), Via Pietro Bucci, 41C, 87036 Rende (CS), Italy, amelio@icar.cnr.it} and  Clara Pizzuti \footnote{National Research Council of Italy (CNR), Institute for High Performance Computing and Networking (ICAR), Via Pietro Bucci, 41C, 87036 Rende (CS), Italy, pizzuti@icar.cnr.it}}
\date{}
\maketitle

\abstract{The roll calls of the Italian Parliament in the XVI legislature are studied by employing multidimensional scaling, hierarchical clustering, and network  analysis. In order to detect changes in voting behavior, the roll calls have been divided in seven periods of six months each. All the methods employed pointed out an increasing  fragmentation of the political parties endorsing the  previous government that culminated in its downfall. By using the concept of modularity at different resolution levels, we identify the community structure of Parliament and its evolution in each of the considered time periods. The analysis performed revealed  as a valuable tool  in detecting trends and drifts of Parliamentarians. It  showed its effectiveness at identifying political parties and at providing insights on the temporal evolution of groups and their cohesiveness, without having at disposal any knowledge  about political membership of Representatives.}

\section{Introduction}

In the last years  political parties in Italy have been affected by a steady fragmentation, with a high number of Parliamentarians leaving  the  group that allowed them to be elected to join another one, often changing party many times.

In this paper we investigate Italian Parliament by using different tools coming from Data Mining and Network Analysis fields with the aim of characterizing the modifications Parliament incurred,  without any knowledge about the ideology or political membership of its Representatives, but relying only on the votes cast by each Parliamentarian.  We consider the roll calls of the period of three years and an half from April 2008 until October 2011, after which there was the fall of the center-right coalition that won the elections. This period has been equally divided in seven semesters and the votes cast by each Parliamentarian have been stored. Note that in our analysis we do not consider the Italian Senate. 

Voting records have been used in two different ways.  In the first approach we directly use them to show party cohesion during the considered period, and apply a multidimensional scaling technique  to reveal political  affinity of Parliamentarians, independently of their true party membership. This kind of analysis is interesting because it is able to reproduce the effective political alliances, without assuming  parties as relevant clusters. 

In the second one, from voting records we compute  similarity between each pairs of Representatives  and try to detect structural organization and evolution of Parliament by applying data mining  and network analysis techniques. In particular, similarity among Parliamentarians is exploited to perform clustering by employing agglomerative hierarchical clustering. The division of Representatives in groups is congruous with that obtained by applying multidimensional scaling, thus showing the robustness of both approaches. 

As regards network analysis techniques, from the similarity matrix  a network is built where nodes correspond to Parliamentarians and an edge between two nodes exists if the similarity between the corresponding Representatives is above  a fixed value. Topological features characterizing the network are studied by computing some well known measurements to quantify structural properties, and  community detection is applied to study the organization of members in groups.    By using the modularity concept  \cite{NewmanGirvan2004}, we identify communities of members that voted similarly, and investigate how the party cohesion evolves along the semesters. The analysis provides an explicit and clear view of the steady fragmentation of the coalition endorsing the center-right government, that caused the majority breakdown. Thus modularity allows a more deep analysis of the internal agreement of parties, and demonstrated a powerful  means to give insights of changes in majority party. 

The investigation of voting records with computational techniques is not new \cite{Jak2004,Pajala04,Porter2007,Zhang2008,WPFMP09,Macon2012}, though this is the first study regarding an Italian institution.

The paper is organized as follows. In the next section we give a brief description of the Italian Parliament organization and the data set used for the analysis. In section \ref{vp}
we describe the voting matrix, compute party cohesion,  and apply multidimensional scaling approach to voting records. In section \ref{ps} the similarity metric used is defined, and the groups obtained by applying hierarchical clustering are showed is section \ref{clus}. Section \ref{net} builds Parliamentarian networks, identifies and visualizes voting record blocks  along the semesters. Section \ref{struct} computes measurements, well known in network analysis, to study the characteristics of Parliamentarian network. Section \ref{modul} investigates community structure. Section \ref{seven} argues about the results obtained for the last semester. Section \ref{rel} gives a description of related work. Section \ref{concl}, finally, concludes the paper and outlines future developments.

\section{Data description}
The \index{Italian Parliament} Italian Parliament of XVI legislature has been elected in April 2008 and it is constituted by 630 representatives originally elected in  5 main political parties: People of Liberty (PDL), League of North (LN),  Democratic Party (PD), Italy of Values (IDV), and Democratic Union of Center (UDC). The majority of center-right that governed Italy until November 2011 was composed by the first two parties. To better understand the analysis we performed, it is important to know that two main events characterized the political organization of Parliament: (1) in July 2010 a group of Representatives divided from PDL  to form a new political party named  Future and Liberty (FL); (2) in December 2010 some Parliamentarians, mainly coming from the center-left coalition, separated from their party to constitute a new coalition, named People and Territory (PT), that endorsed the center-right government, allowing it to rule the country for other almost ten months. Furthermore, along all the three years and an half, several Representatives abandoned their party to move in a group called Mixed.
The Italian Parliament maintains a database of the legislative activity by storing, for each bill voted, the list of votes cast by each Representative. From the web site  http://parlamento.openpolis.it it is possible to download the voting record of each Parliamentarian, together with some personal information, such as territorial origin, and actual group membership. For every roll call, the \index{Openpolis database} Openpolis database stores the vote of each Parliamentarian in three ways: ``yes", ``no", and ``not voting". This last kind of vote can be due to either absence or abstention, but they are treated in the same manner. 
\begin{table}[t]
\caption{Number of voted measures for each semester.}
\label{tabsem}
\centering
\begin{tabular}{p{1cm}p{1cm}p{1cm}p{1cm}p{1cm}p{1cm}p{1cm}}
\hline
I  & II  & III  & IV  & V & VI  & VII\\
\hline
386 & 422 & 328 & 343 & 373 & 332 & 89\\
\hline
\end{tabular}
\end{table}

\section{Analysis of voting patterns} \label{vp}

We collected the roll calls of the Italian Parliament in the period starting from April 2008 until October 2011, after which there was the fall of the center-right coalition that won the elections. This period of three years and an half has been equally divided in seven semesters and the votes cast by each Parliamentarian have been stored in matrices of size $n \times m$, where $n$ is the number of Parliamentarians, and $m$ is the number of bills voted in the reference period. Since some Parliamentarians, for several reasons, never voted, they have been eliminated. Thus the number $n$ of Representatives reduced to 612. As regards $m$, it  assumes a different value, depending on the semester. The number of bills voted is reported in Table \ref{tabsem}.  
Seven voting matrices have been built in the following way: an element $A_{ij}$ of a voting matrix $A$  is +1 if the Representative $i$ voted ``yes" on measure $j$, -1 if he voted ``no", and $0$ if he did not vote. The \index{voting matrices} voting matrices are exploited to study the voting behavior of the Italian Parliament in two different ways. In the first approach we use them to compute party cohesion and to characterize the political  affinity of Parliamentarians, independently of their true party membership. In the second one, we compute  similarity for each pairs of Representatives and try to detect structural organization and evolution  by applying hierarchical clustering and community detection based on the concept of modularity.    
 \subsection{Party Cohesion}
Given the voting matrices, the first investigation that can be done is to compute the cohesion of each  political party along the considered period and compare the results obtained. To this end, the \index{agreement index} \emph{agreement index} \cite{Hix2005} measures the level of cohesion within a party by exploiting the number of equal votes for each roll call. The agreement index for each roll call is defined as follows: 
\begin{equation}\label{}
AI_i=\frac{ max\{y_i,n_i,a_i\}-\frac{y_i+n_i+a_i-max\{y_i,n_i,a_i\}}{2}}{y_i+n_i+a_i}
\end{equation}

where $y_i$ is the number of members who voted ``yes" in the voting $i$,
$n_i$ is the number of members who voted ``no", and $a_i$ is the number of members who did not vote.
\index{Group cohesion} Group cohesion is then computed as the average of agreement indices for all the roll calls: \begin{equation}\label{}
AI=\frac{\sum^m_i AI_i}{m}
\end{equation} 
The agreement index ranges from 0 (complete disagreement) to 1 (complete agreement). 

\begin{figure}[h]
\includegraphics[scale=.50]{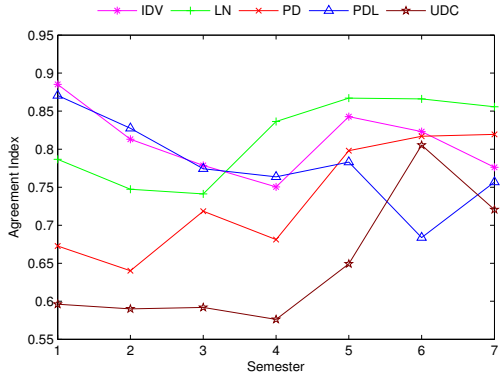}
\caption{Agreement index of parties for all the semesters.}
\label{agree}      
\end{figure}

Figure \ref{agree} displays the trend of agreement index of the 5 main political parties during the seven semesters. It is clear from the figure that the opposition parties show an increasing cohesion, while PDL, that started with a value near to 0.9, has a constant downtrend until the sixth semester,  with a slight increment in the last semester. The variation of internal cohesion well reflects the actual political situation along the considered periods. 

\begin{figure}[th!]
  \centering
  \subfigure[I Semester: April-September 2008]
  {\includegraphics[width=0.48\textwidth]{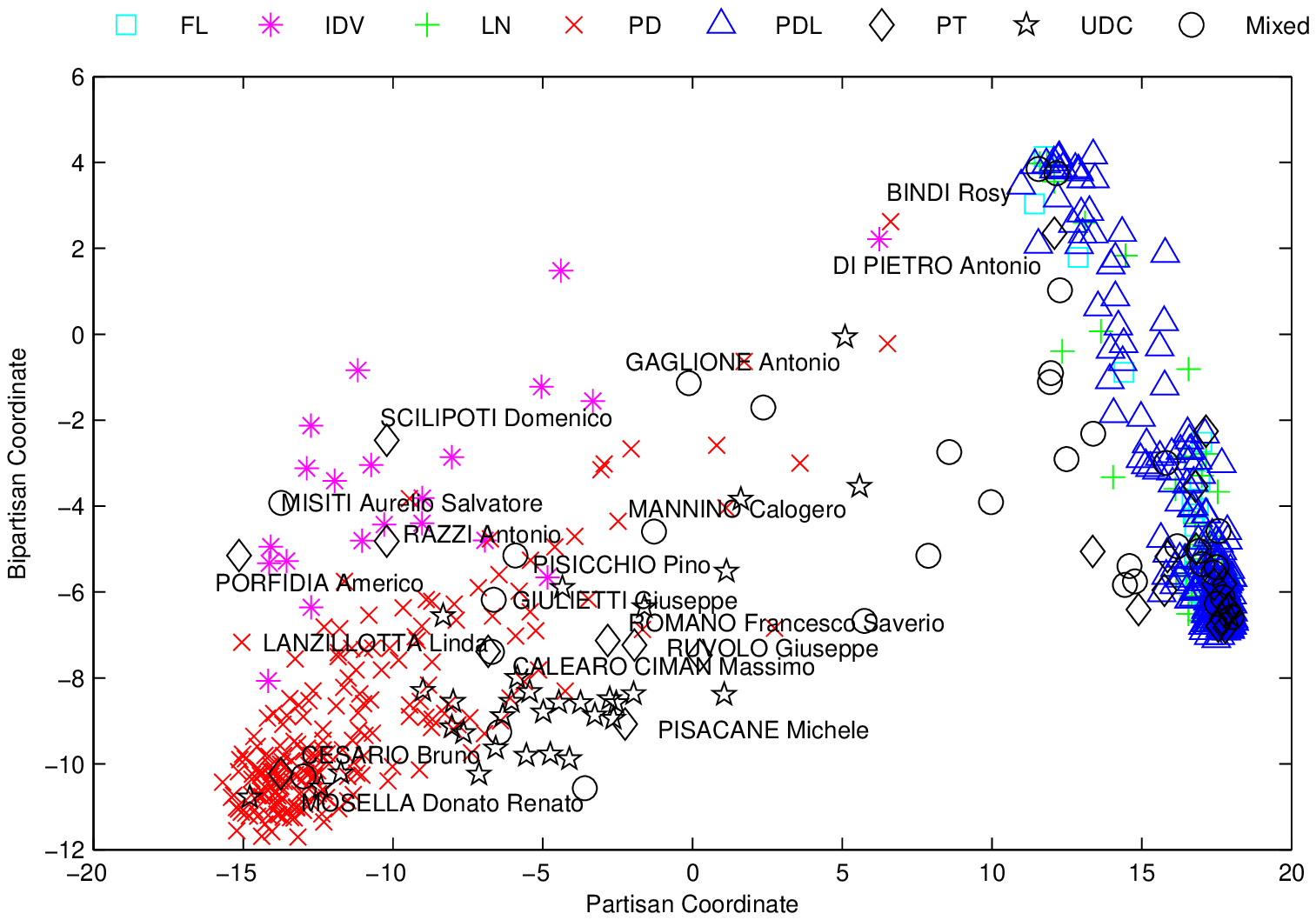}}
  \subfigure[II Semester:  October 2008-March 2009 \label{}]
  {\includegraphics[width=0.48\textwidth]{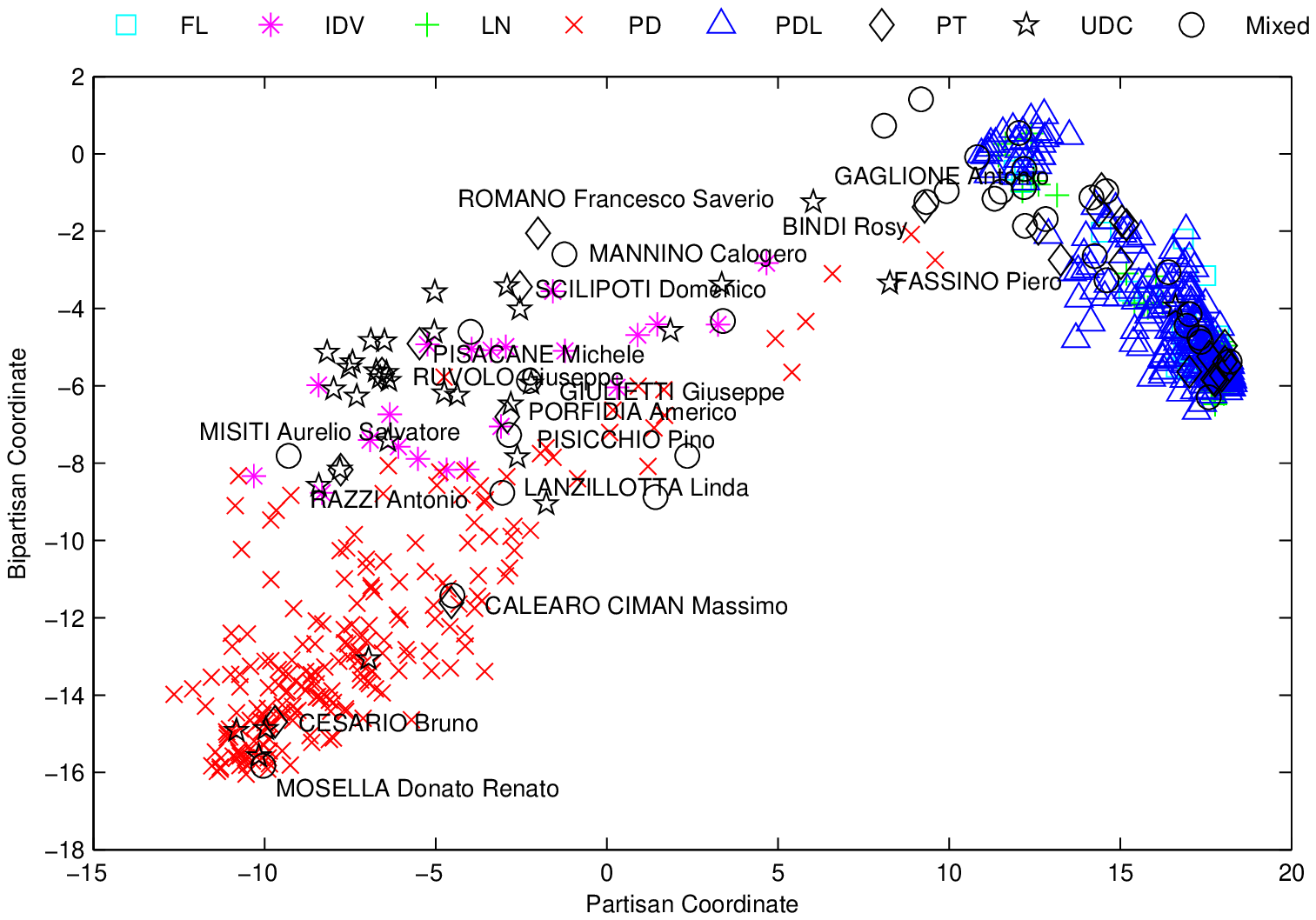}}
  \subfigure[III Semester: April-September 2009 \label{}]
  {\includegraphics[width=0.48\textwidth]{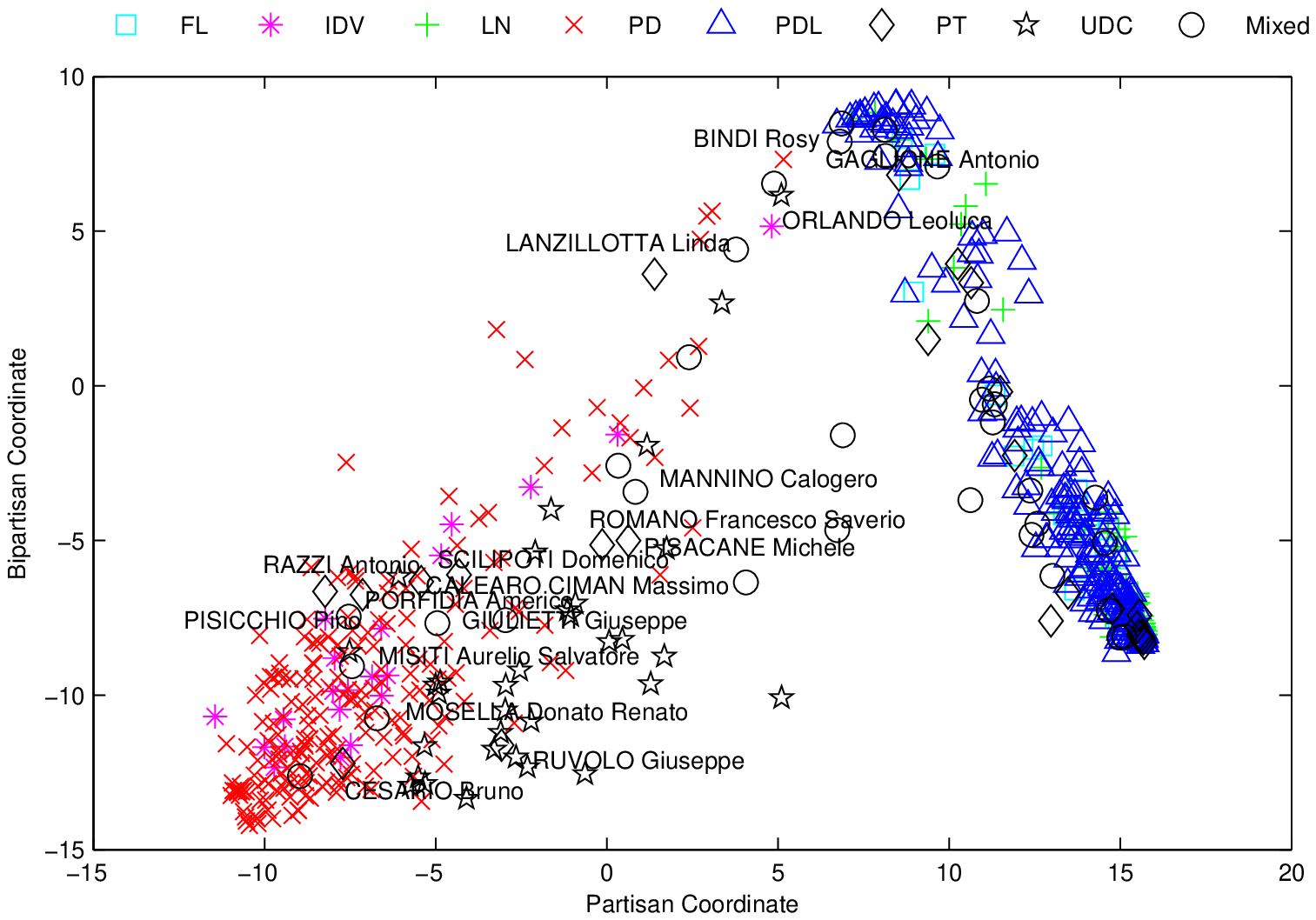}}
  \subfigure[IV Semester: October 2009-March 2010\label{}]
  {\includegraphics[width=0.48\textwidth]{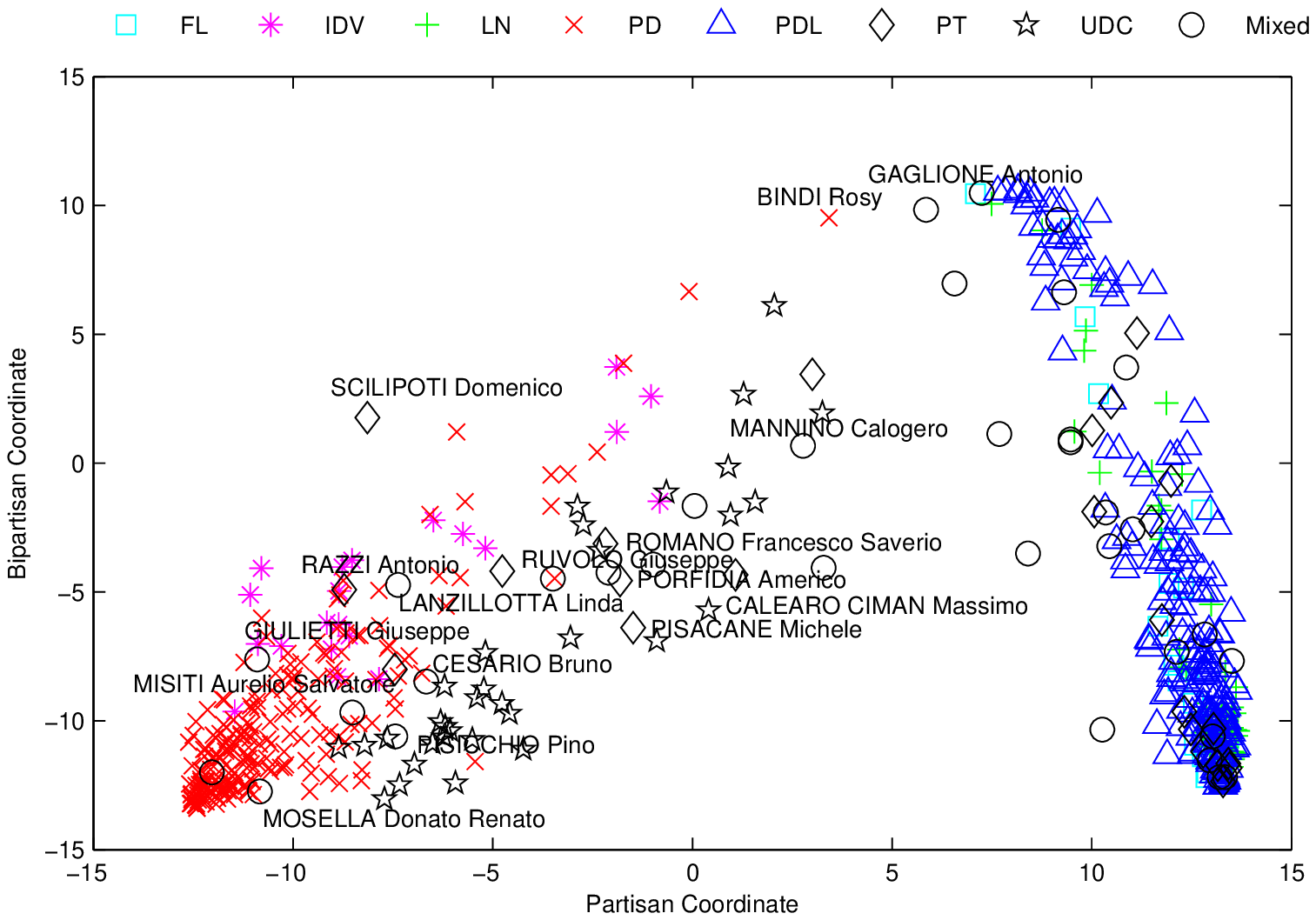}}
\subfigure[V Semester: April-September 2010 \label{}]
  {\includegraphics[width=0.48\textwidth]{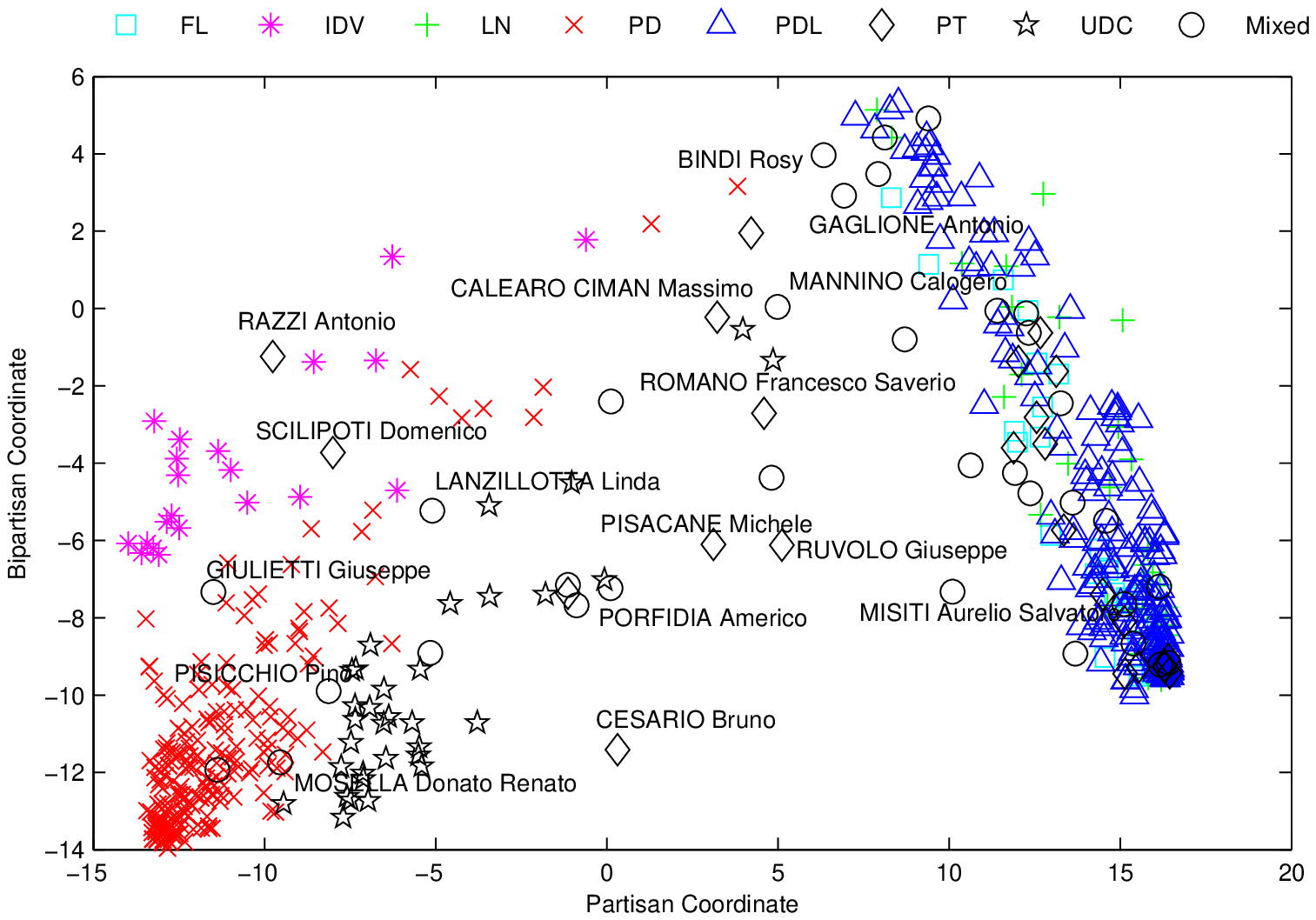}}
  \subfigure[VI Semester: October 2010-March 2011 \label{}]
  {\includegraphics[width=0.48\textwidth]{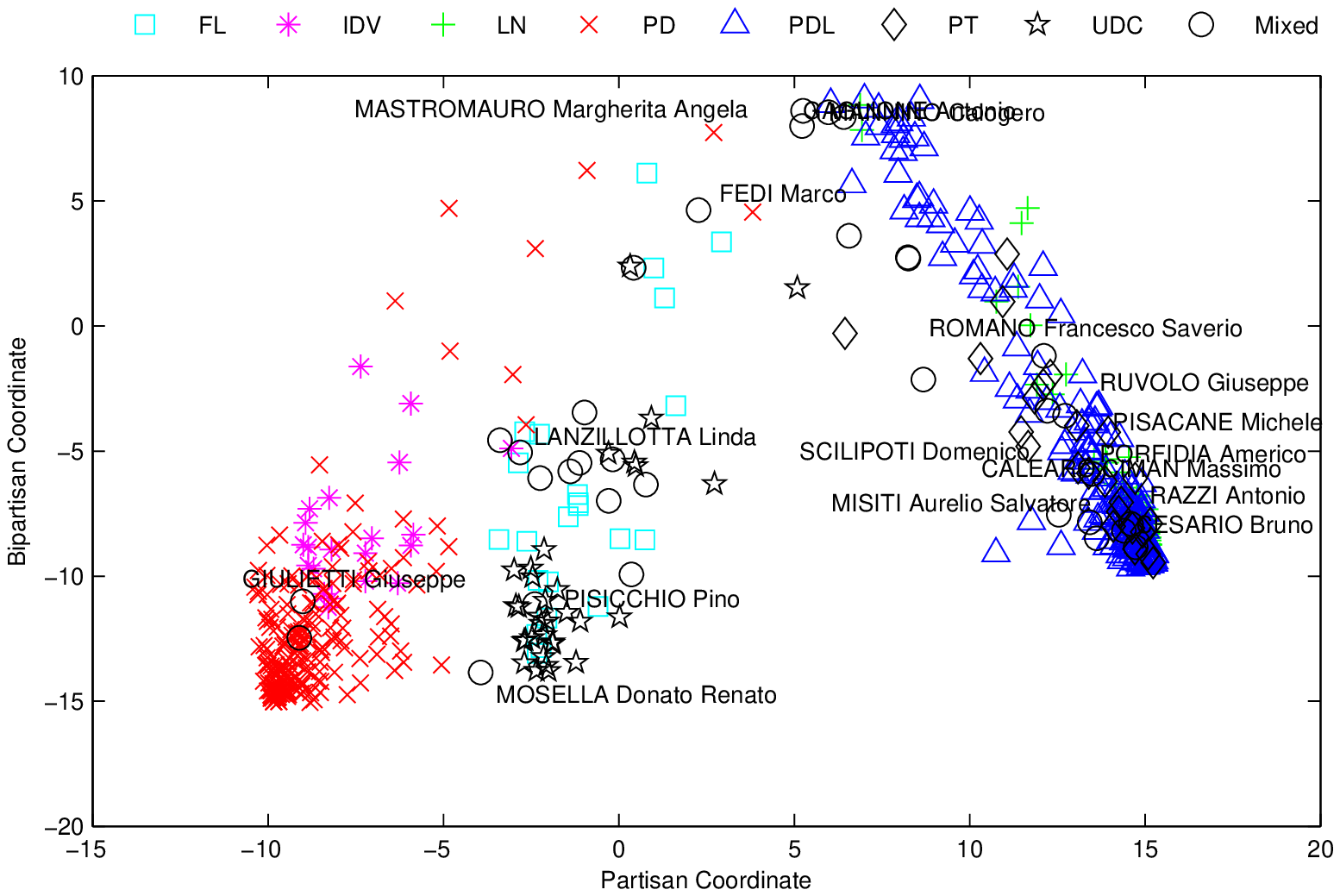}}

  \caption{Singular value decomposition of the Italian Parliament voting behavior for each of the six semesters starting from April 2008 until March 2011.}
\label{svd}
\end{figure}

\subsection{Singular Value Decomposition}
We now analyze the voting behavior of Italian Parliament by applying the well known \index{multidimensional scaling} multidimensional scaling technique \emph{Singular Value Decomposition} ($SVD$) \cite{Strang2005}, whose advantages with respect to other techniques have been discussed in \cite{Brazill2002}. 
Let $A$ be an $n \times m$ voting matrix where rows correspond to  Representatives and columns to the votes cast to approve a law. The \index{Singular Value Decomposition} {\em Singular Value Decomposition} of $A$ is any factorization of the form 

\begin{equation}\label{}
A = U \times \Lambda \times V^T
\end{equation}
where $U$ is an $n \times n$ orthogonal matrix, $V$ is an $m \times m$ orthogonal matrix and $\Lambda$ is an $n \times m$ diagonal matrix with $\lambda_{ij}=0$ if $i\neq j$. The diagonal elements $\lambda_i$ are called the \emph{singular values} of $A$. 
It has been shown that there exist matrices $U$ and $V$ such that the diagonal elements of $\Lambda$ are the square roots of the nonzero eigenvalues of either $A A^T$ or $A^T A$, and they can be  sorted such that $\lambda_1 \geq \lambda_2 \geq \ldots \geq \lambda_m$ \cite{Strang2005}.  Geometrically this factorization defines a rotation of the axis of the vector space defined by $A$ where $V$ gives the directions, $\Lambda$ the strengths of the dimensions, and $U \times \Lambda$ the position of  the points along the new axis.
Intuitively, the $U$ matrix can be viewed as a similarity matrix among the rows of $A$, i.e. the Representatives, the $V$ matrix as a similarity matrix among the columns of $A$, i.e. the votes cast for each law, the $\Lambda$ matrix gives a measure of how much
the data distribution is kept in the new space \cite{jain-88}. If the singular values $\lambda_i$ present a fast decay, then  $U \times \Lambda$ provides a good approximation of the original voting matrix $A$. In particular, by projecting on the first two coordinates, we obtain a compressed representation of the voting matrix that approximates it at the best. The visualization of the projected approximation matrix, allows to identify groups of Representatives that voted in a similar way on many bills. As observed in \cite{Porter2007}, the first coordinate correlates to party membership, thus it is called the \emph{partisan} coordinate \index{partisan coordinate}. The second coordinate correlates to how often a Representative voted with the majority, thus it is called the \emph{bipartisan} coordinate \index{bipartisan coordinate}. 

Figure \ref{svd} shows the application of $SVD$ on the voting records of the Italian Parliament for the first six semesters of the current legislature. Each point corresponds to the projection of votes cast by a single Parliamentarian onto the leading two eigenvectors partisan and bipartisan. Each party has been assigned a different color and symbol.  The main objective of this analysis was to study the changes in voting behavior of those Parliamentarians that moved from the opposition coalition to the majority one. Thus we selected some members of PT party and Mixed group, and visualized their names on all the figures. First of all we point out that the representation of the two coalitions center-right and center-left, and their evolution along the three years, summarized by the six figures, is very impressive.  

Figure \ref{svd}(a) clearly shows a compact center-right aggregation,  a less cohesive, but clearly distinguishable,  center-left alliance, and a strong connected PD sub-group (left bottom). It is worth to note that this sub-group maintains its connectedness for all the time periods,  with a slight dispersion in the second semester. The same cohesiveness is shown by PDL and LN, as expected. Moreover FL, which was included in PDL until July 2010, demonstrated its  political disagreement in the sixth semester by coming nearer  to UDC, as effectively happened. 
As regards the chosen members of PT and Mixed group,  we can observe a steady movement from the center-left coalition to the center-right one since the fourth semester. This shift is much more evident in the 5th semester, when the voting behavior of these Representatives approached closer and closer to center-right majority. In fact, all the Parliamentarians located in the central part of Figure \ref{svd}(e), appear at right in Figure \ref{svd}(f), indistinguishable from the majority coalition.   

We also notice that there is a PD Parliamentarian positioned upper, near the right coalition, for  five semesters. Because of the interpretation of the bipartisan coordinate, her location means that she mostly voted with the majority. This dissimilarity from the own political party, perhaps can be  explained by the fact that this Representative was vice-president of the chamber.

Analysis of voting behavior with Singular Value Decomposition is thus a powerful tool to characterize political ideology of Parliamentarians, and to trace the evolution of their position
along consecutive time periods. $SVD$ is able to find structural patterns and latent information in the voting records without any knowledge about the political orientation of Representatives.  

\section{Parliamentarians similarity} \label{ps}
There can be different ways of defining similarity between two Parliamentarians from the voting matrix. For example, Jakulin and Buntine \cite{Jak2004} used the  mutual information concept. However, as observed by the authors, if two members always vote in the opposite way, they also are considered similar. We think that this kind of proximity measure misrepresents the Representative closeness, thus we employed a more suitable measure. Considering that when two Representatives cast a vote the values ``yes" and ``no" should be considered equally important in comparing their political affinity, we adopted the proximity measure known as \emph{simple matching coefficient} \index{simple matching coefficient} ($SMC$) \cite{tan06}.   We ignored the cases when at least one of the two did not vote because, as already pointed out, this means either abstention or absence, and we cannot distinguish between them. Thus there can be four different outcomes: (1) $yy$, both voted ``yes", (2) $nn$, both voted ``no", (3) $yn$, the first Parliamentarian voted ``yes" and the second one ``no", (4) $ny$, the first Parliamentarian voted ``no" and the second one ``yes". Then the $SMC$ of Parliamentarians $p_1$ and $p_2$ is defined as 
\begin{equation}\label{}
SMC(p_1,p_2)=\frac{yy + nn}{yy+ nn + yn + ny}
\end{equation} 
The simple matching coefficient thus computes  the fraction of equal votes, both ``yes" or ``no", with respect to the total votes they cast.    
The similarity metric defined allows us to measure the closeness of each pair of Parliamentarians on the base of their voting behavior. In such a way a symmetric similarity matrix $M$ among all the Parliamentarians can be built, and their proximity with the members of the same or opposite parties  studied. A summarized view of the affinity between each couple of Representatives can be done in different ways. In the following we first apply a hierarchical clustering algorithm, and then we give a graphical representation of the similarity matrix.

\begin{figure}[ht!]
  \centering
  \subfigure[I Semester \label{}]
  {\includegraphics[width=0.48\textwidth]{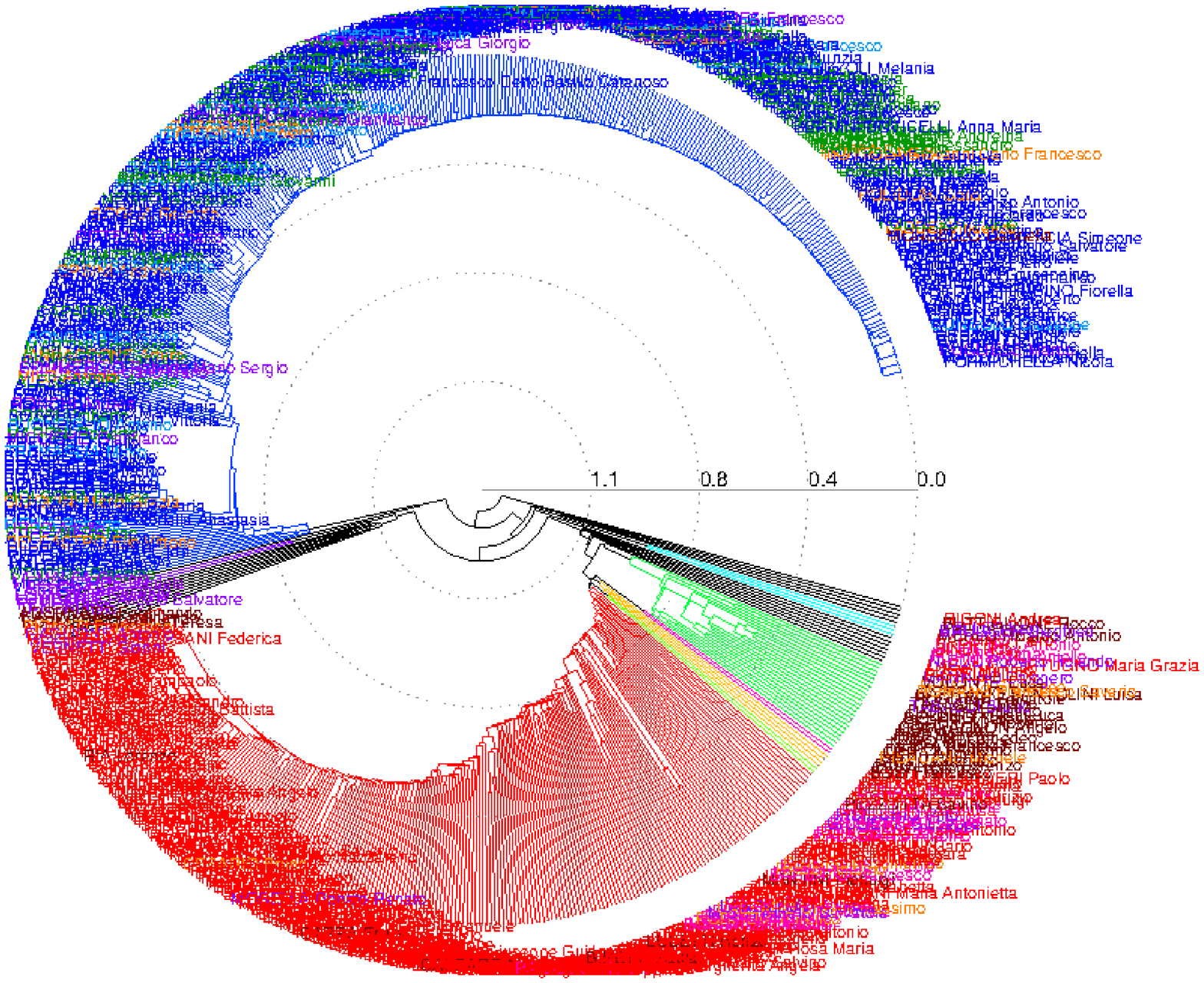}}
  \subfigure[II Semester \label{}]
  {\includegraphics[width=0.48\textwidth]{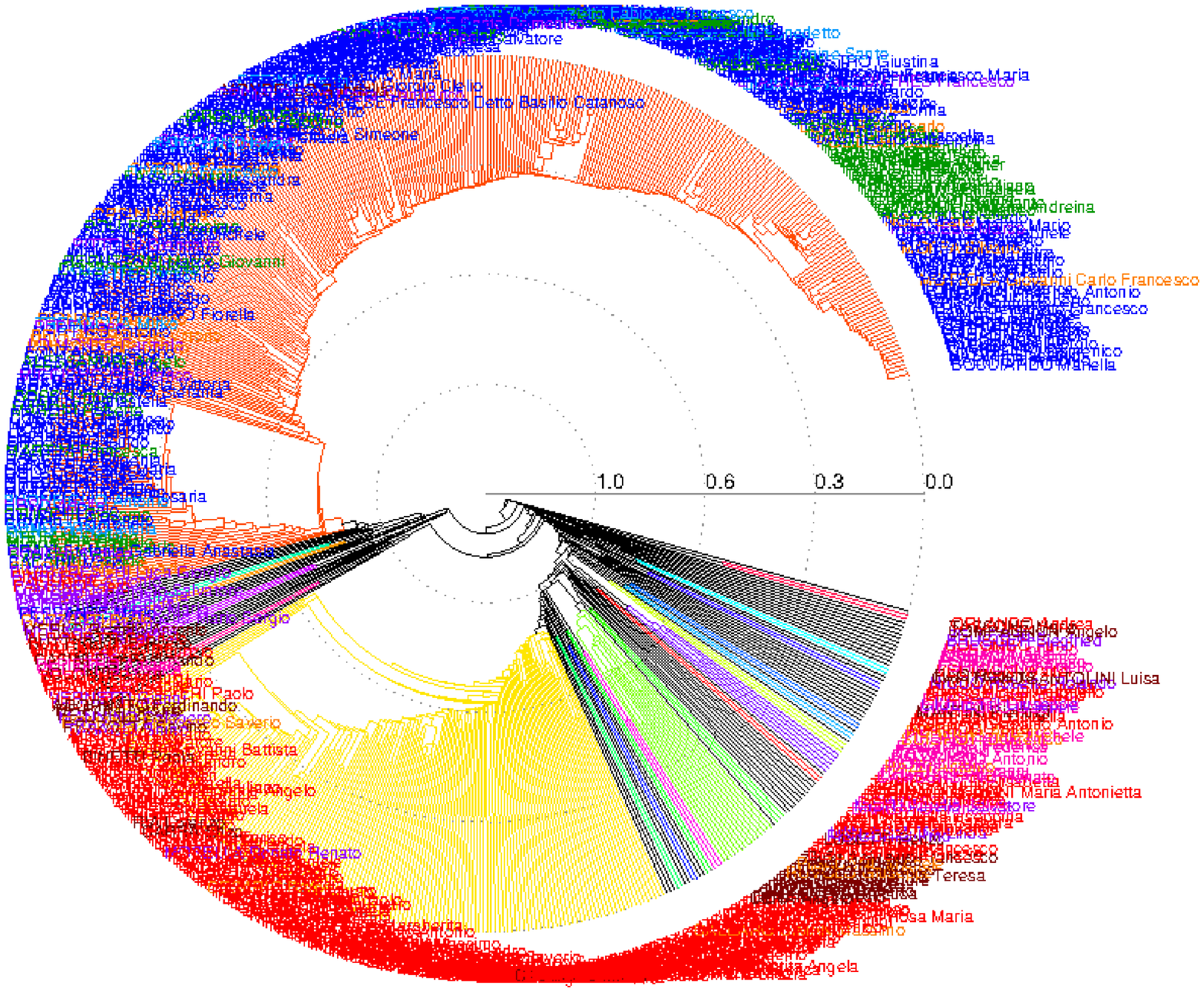}}
  \subfigure[III Semester \label{}]
  {\includegraphics[width=0.48\textwidth]{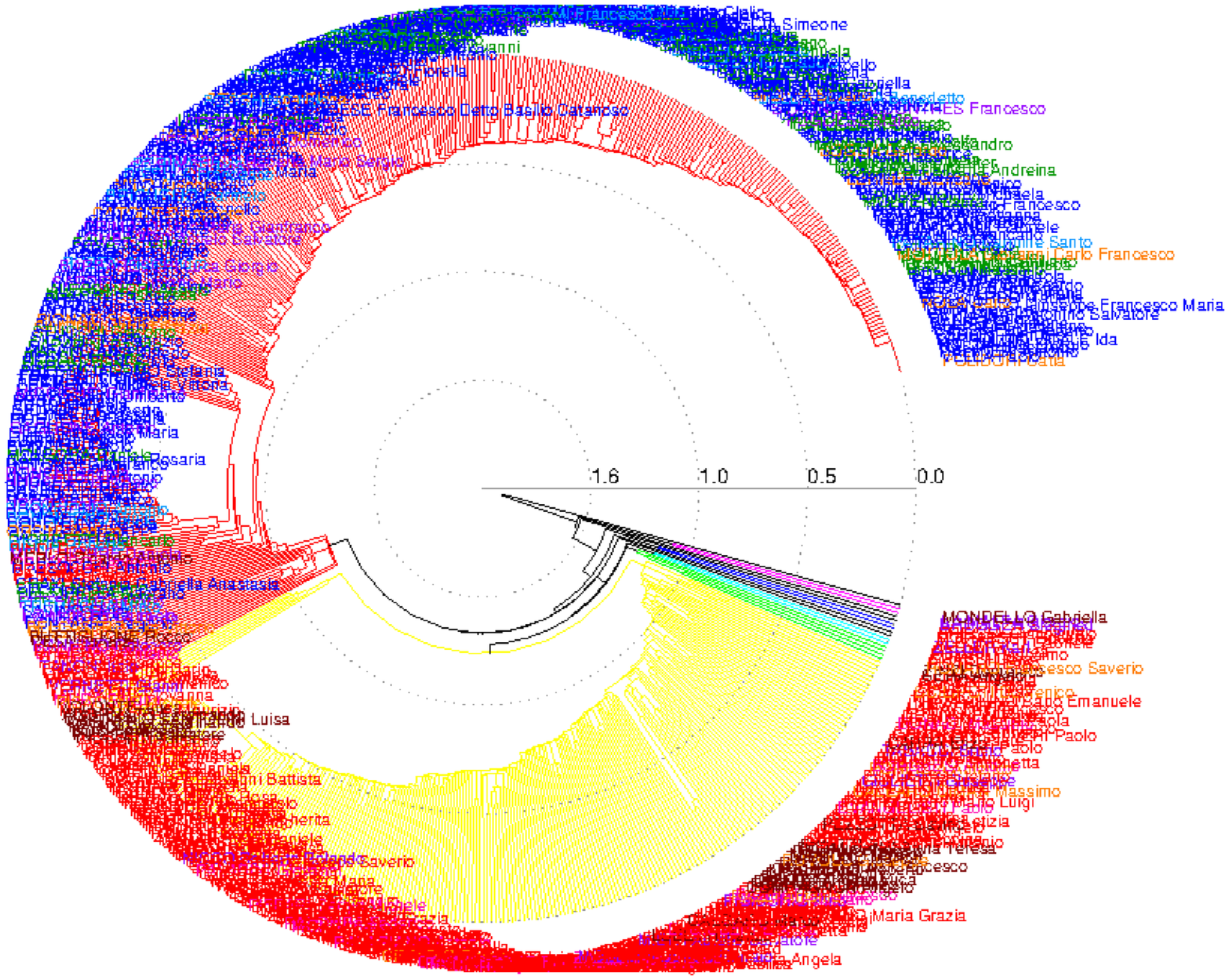}}
  \subfigure[IV Semester \label{}]
  {\includegraphics[width=0.48\textwidth]{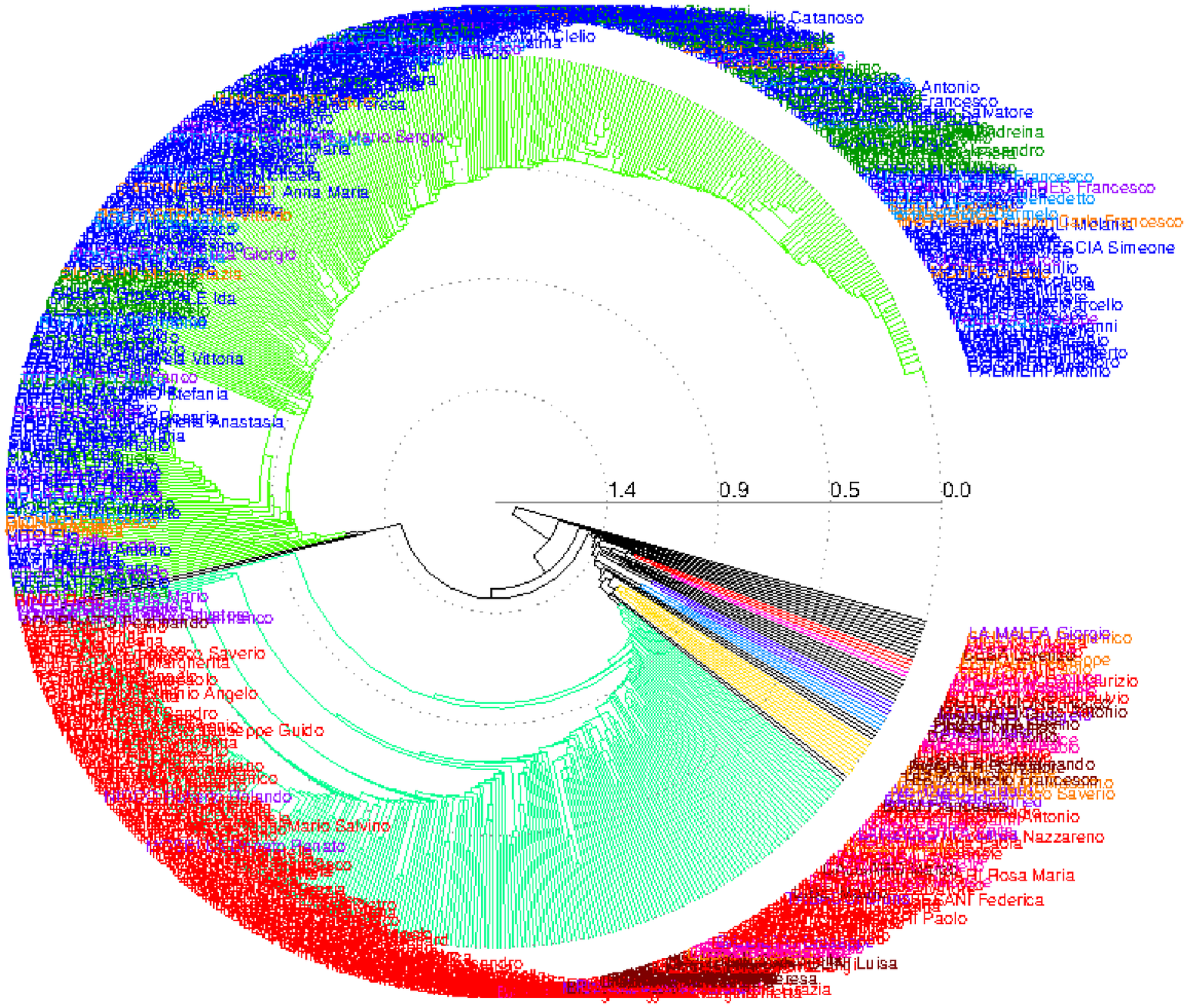}}
\subfigure[V Semester \label{}]
  {\includegraphics[width=0.48\textwidth]{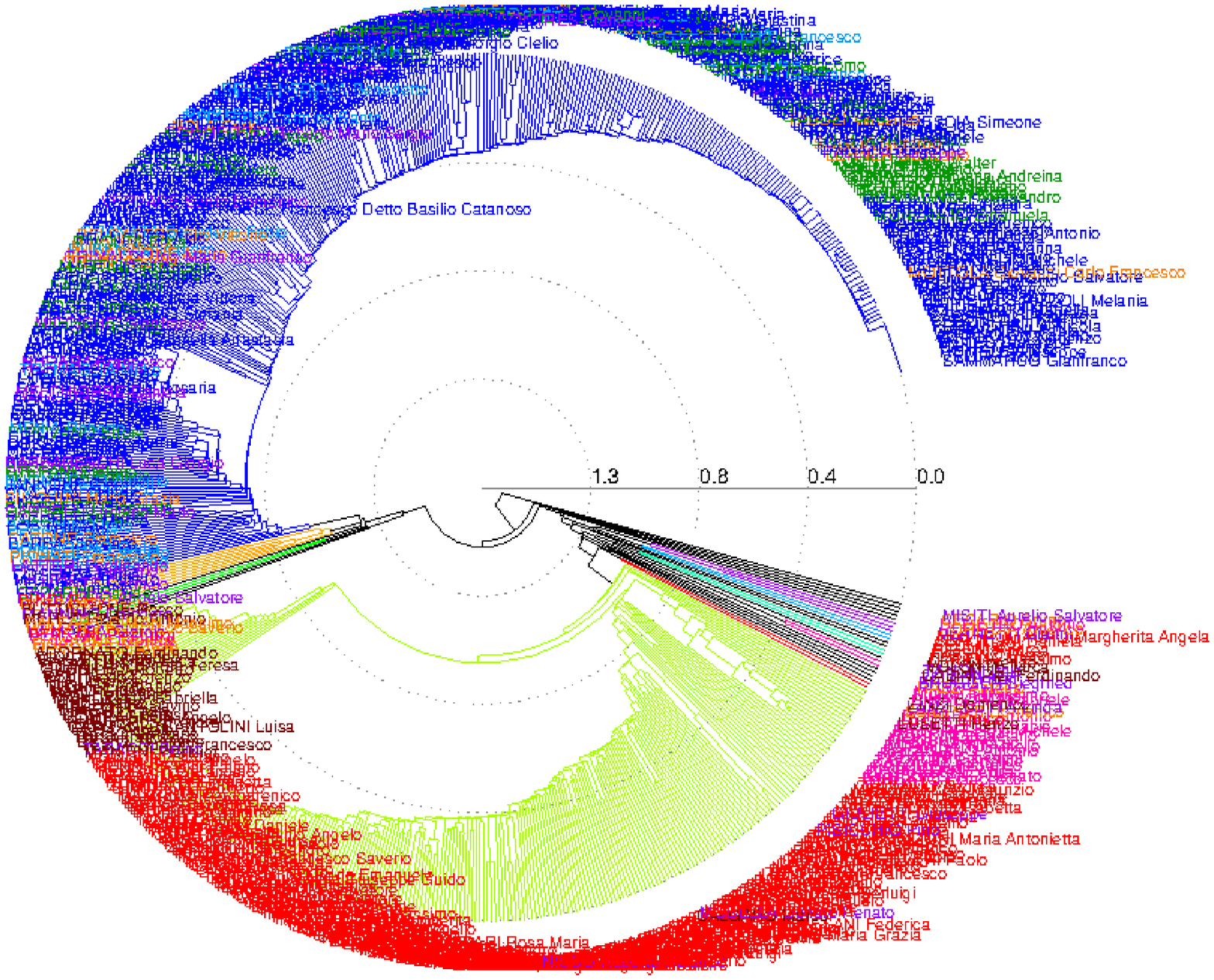}}
  \subfigure[VI Semester \label{}]
  {\includegraphics[width=0.48\textwidth]{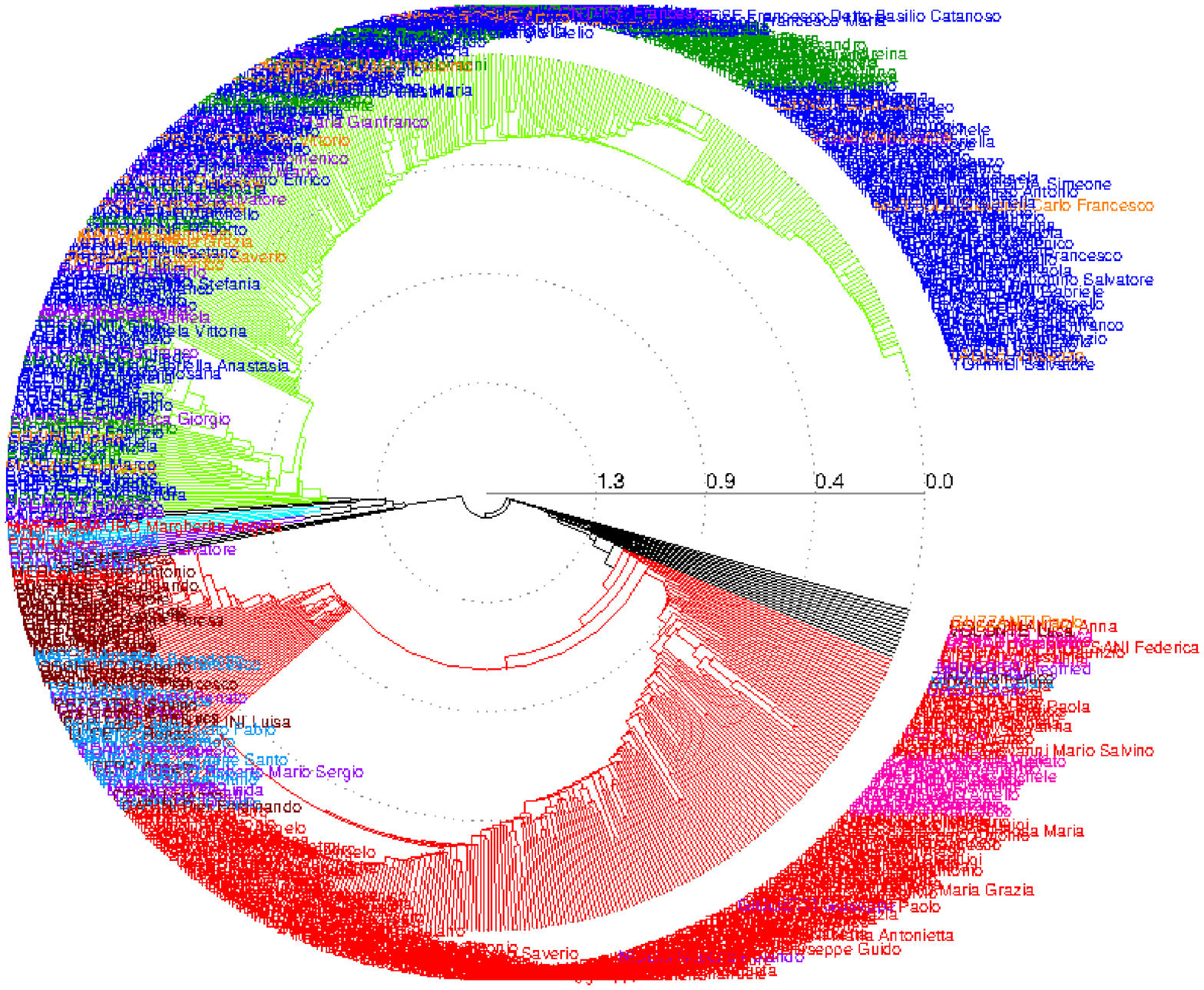}}

  \caption{Dendrograms obtained by the single linkage clustering algorithm for each semester. Internal colors correspond to the clusters found by the algorithm, external colors to the true parties. The association color-party is the following: FL: cyan, IDV: magenta, LN: green, PD: red, PDL: blue, PT: orange, UDC: brown, Mixed: violet.}
\label{dendr}
\end{figure}

\section{Hierarchical clustering} \label{clus}
We apply the agglomerative hierarchical clustering \index{hierarchical clustering} method known as  \emph{single linkage clustering} \index{single linkage clustering} \cite{tan06}. The algorithm uses the smallest distance between two Parliamentarians and it generates a  hierarchical cluster tree known as \emph{dendrogram} \index{dendrogram}.  The dendrogram shows the cluster-subcluster relationships and the hierarchical structure of the merged groups. 
Figure \ref{dendr} represents very well the political alliances along all the semesters\footnote{Enlarged figures of all the dendrograms can be downloaded from https://sites.google.com/site/alessiaamelio/software-tools}. 
The colors inside the dendrogram represent the clusters found by the algorithm. Attached to the leaves there are the names of the corresponding politicians,
painted with the colors of the true associated party.\\
In Figure \ref{dendr}(a)  we can observe as the two main political parties, PD in red and PDL in blue, correspond to the two main clusters of the dendrogram for all the semesters. The other parties
(IDV in magenta, FL in cyan, LN in green, PT in orange, UDC in brown, and Mixed in violet) are clusters of smaller size, or they are merged inside the main clusters.
For example, LN party is grouped together with PDL in all the semesters, reflecting the real political (center-right) alliance between PDL and LN. Another similar case is IDV: most of the members are grouped with the PD while some of them appear in different clusters for all the semesters.\\
Let us now consider the remaining parties.  FL, as already described, was included into PDL until July 2010, when internal problems caused the movement of FL in the direction of center-left alliance. 
This phenomenon is captured from the clustering process. In fact FL is included into the majority for the semesters I-V (Figures \ref{dendr}(a-e)), while in the 6th semester all the members of FL are separated from PDL and grouped together with the opposite part (Figure \ref{dendr}(f)).\\
In order to analyze more clearly the trend of PT and Mixed parties, we looked not only at the dendrograms  but also at the confusion matrices generated for all the semesters.
They show what really happened along the semesters of the  legislature: the gradual movement of PT and of some members of the Mixed group in the direction of the center-right alliance.\\
Furthermore, it is interesting to observe that UDC is recognized from the clustering process as a group (Figure \ref{dendr}(a)), while in the 6th semester (Figure \ref{dendr} (f)) it appears  together with FL and grouped with PD. This is due to the political alliance between the UDC and FL and to the movement of both parties in the direction of the center-left alliance.\\
It is worth to note as the main voting patterns \index{voting patterns} revealed by hierarchical clustering totally agree with the results of the $SVD$ analysis  performed in the previous section.

\begin{figure}[ht!]
  \centering
  \subfigure[I Semester\label{}]
  {\includegraphics[width=0.48\textwidth]{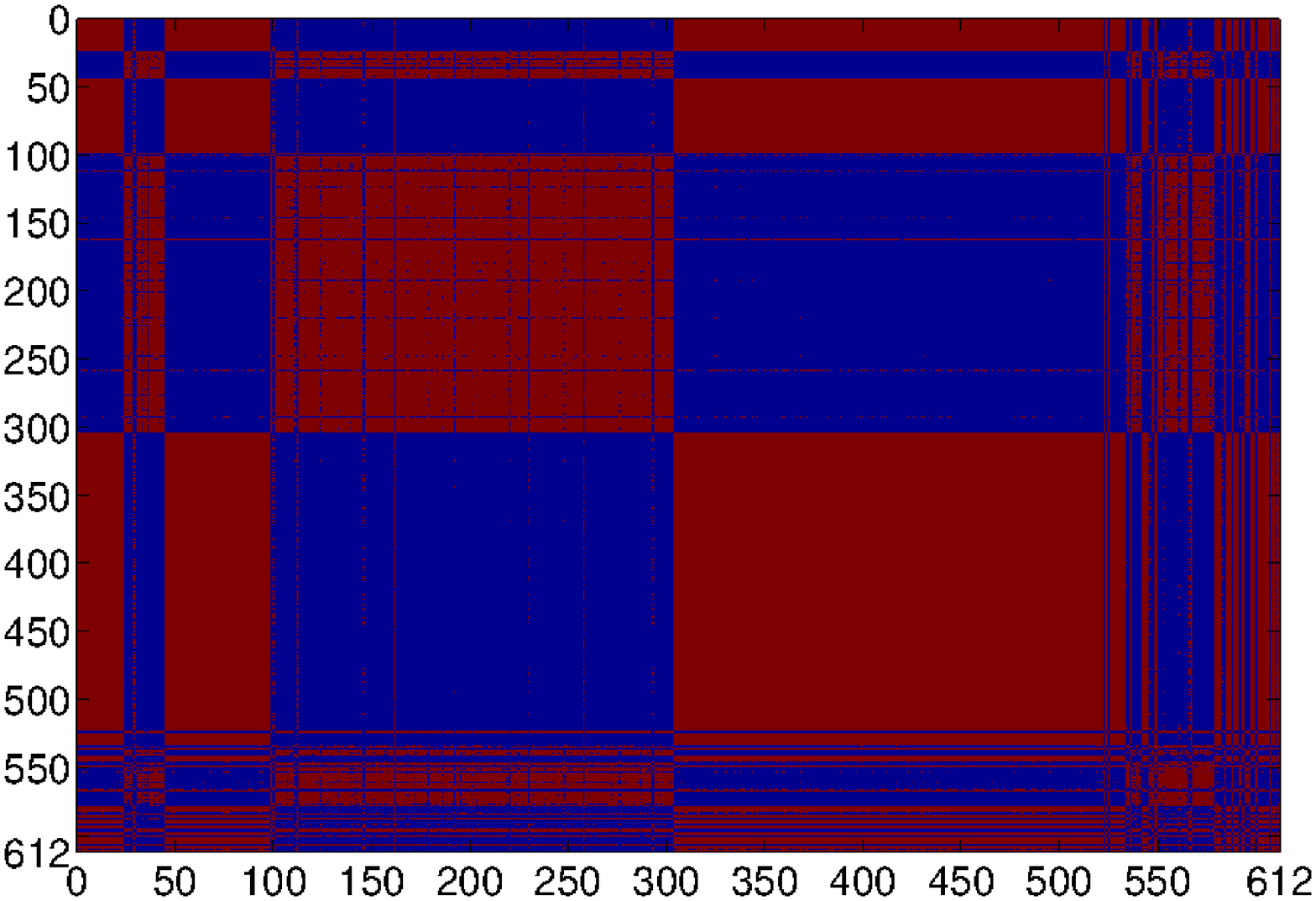}}
  \subfigure[II Semester \label{}]
  {\includegraphics[width=0.48\textwidth]{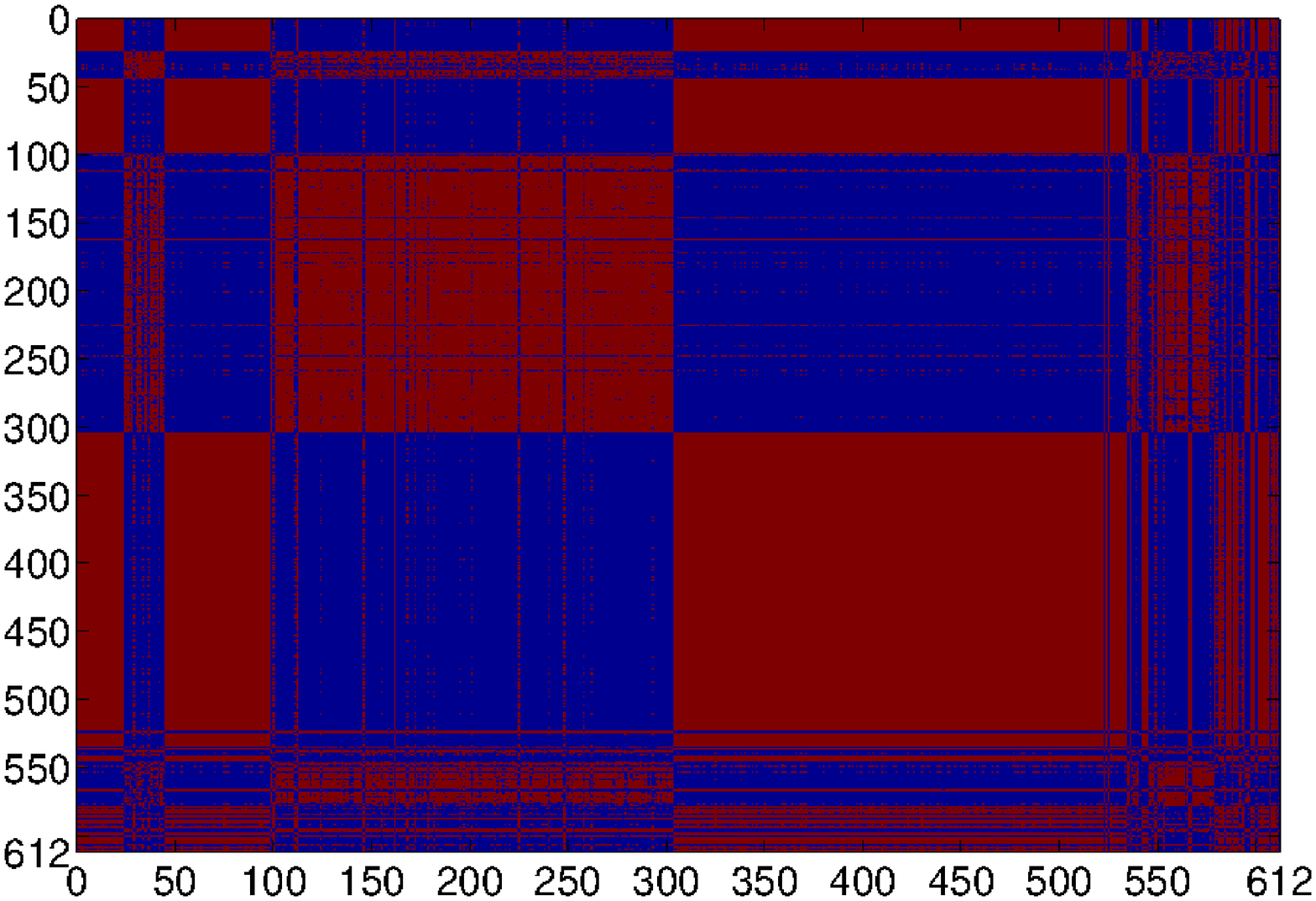}}
  \subfigure[III Semester \label{}]
  {\includegraphics[width=0.48\textwidth]{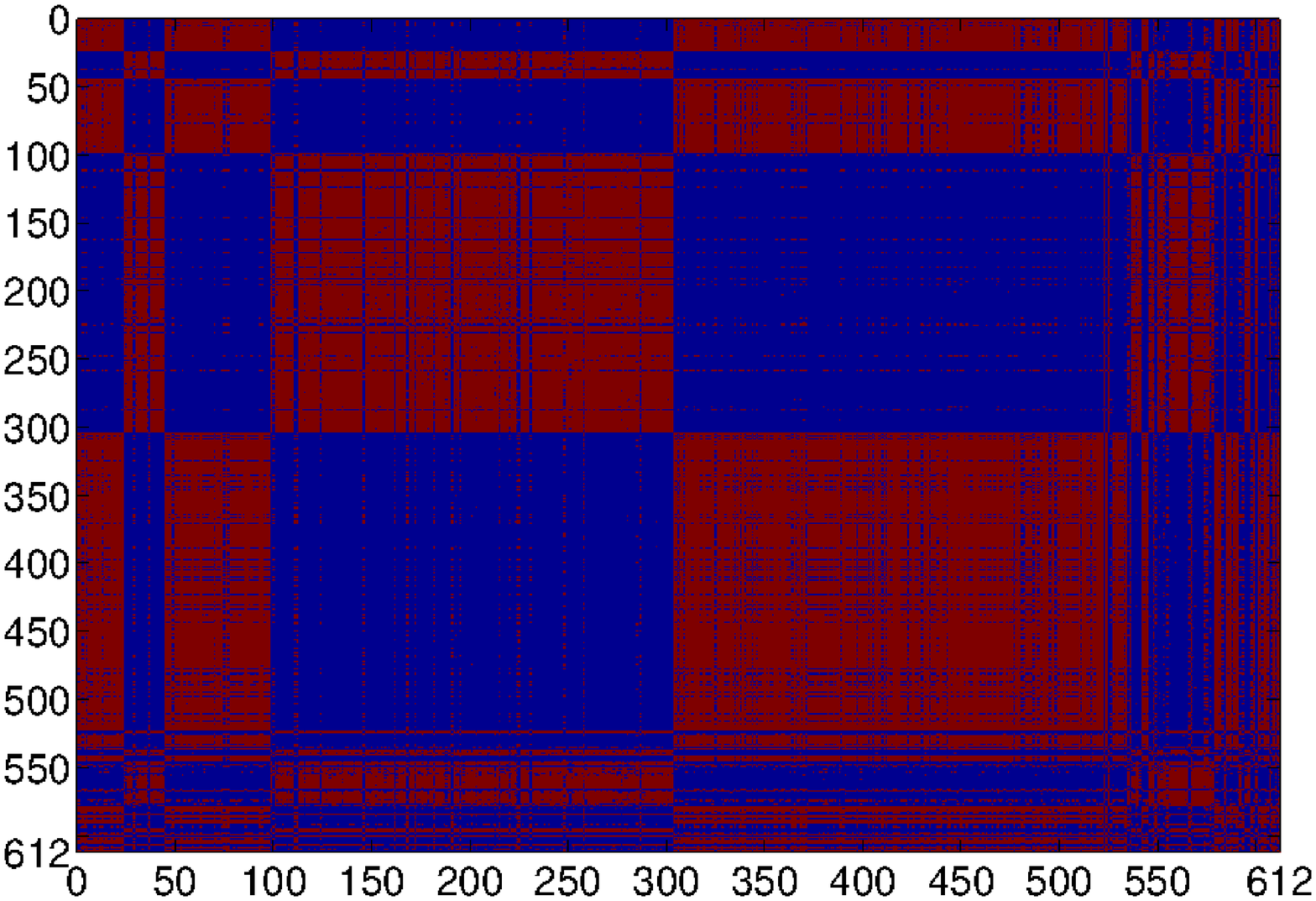}}
  \subfigure[IV Semester\label{}]
  {\includegraphics[width=0.48\textwidth]{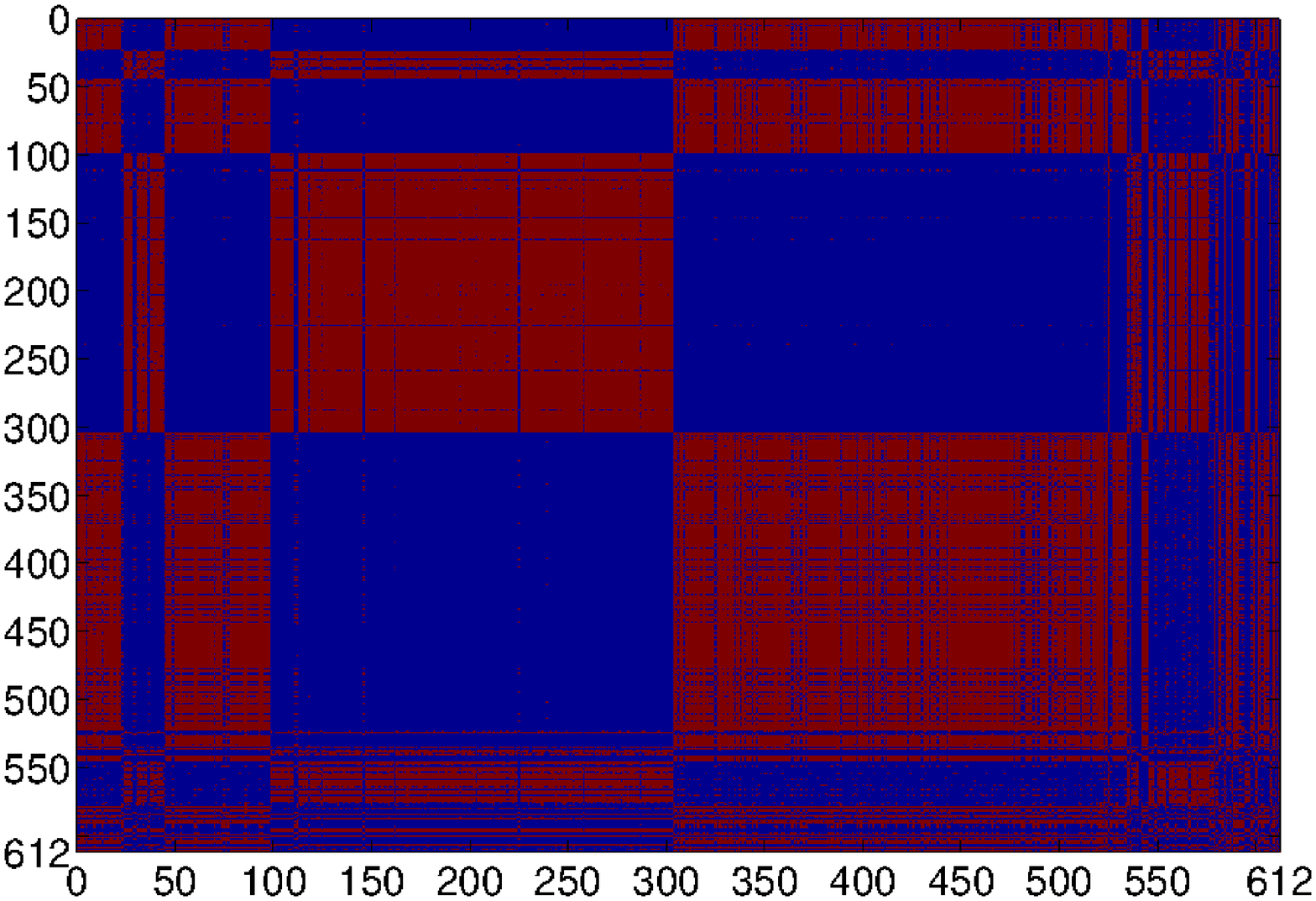}}
\subfigure[V Semester \label{}]
  {\includegraphics[width=0.48\textwidth]{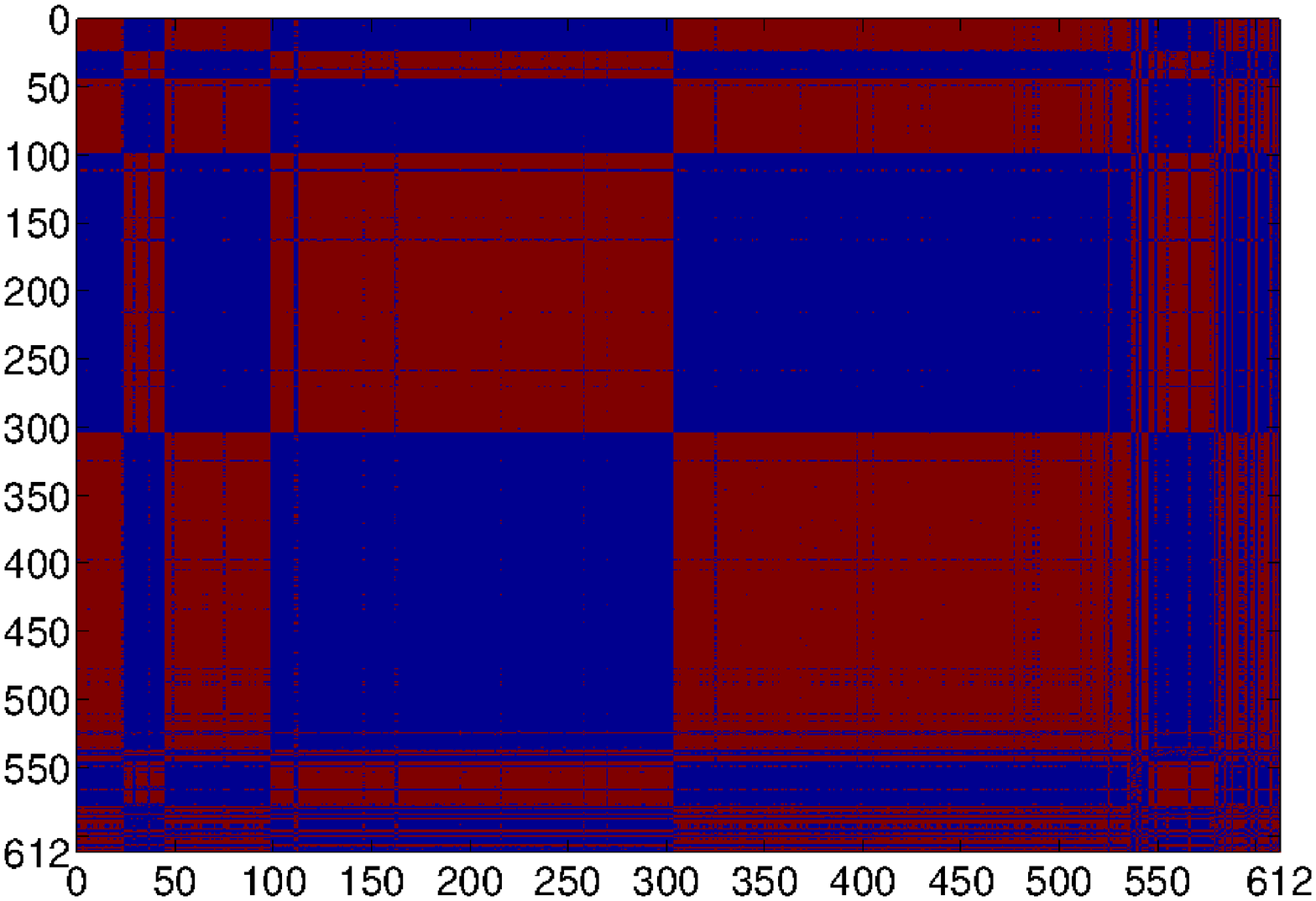}}
  \subfigure[VI Semester \label{}]
  {\includegraphics[width=0.48\textwidth]{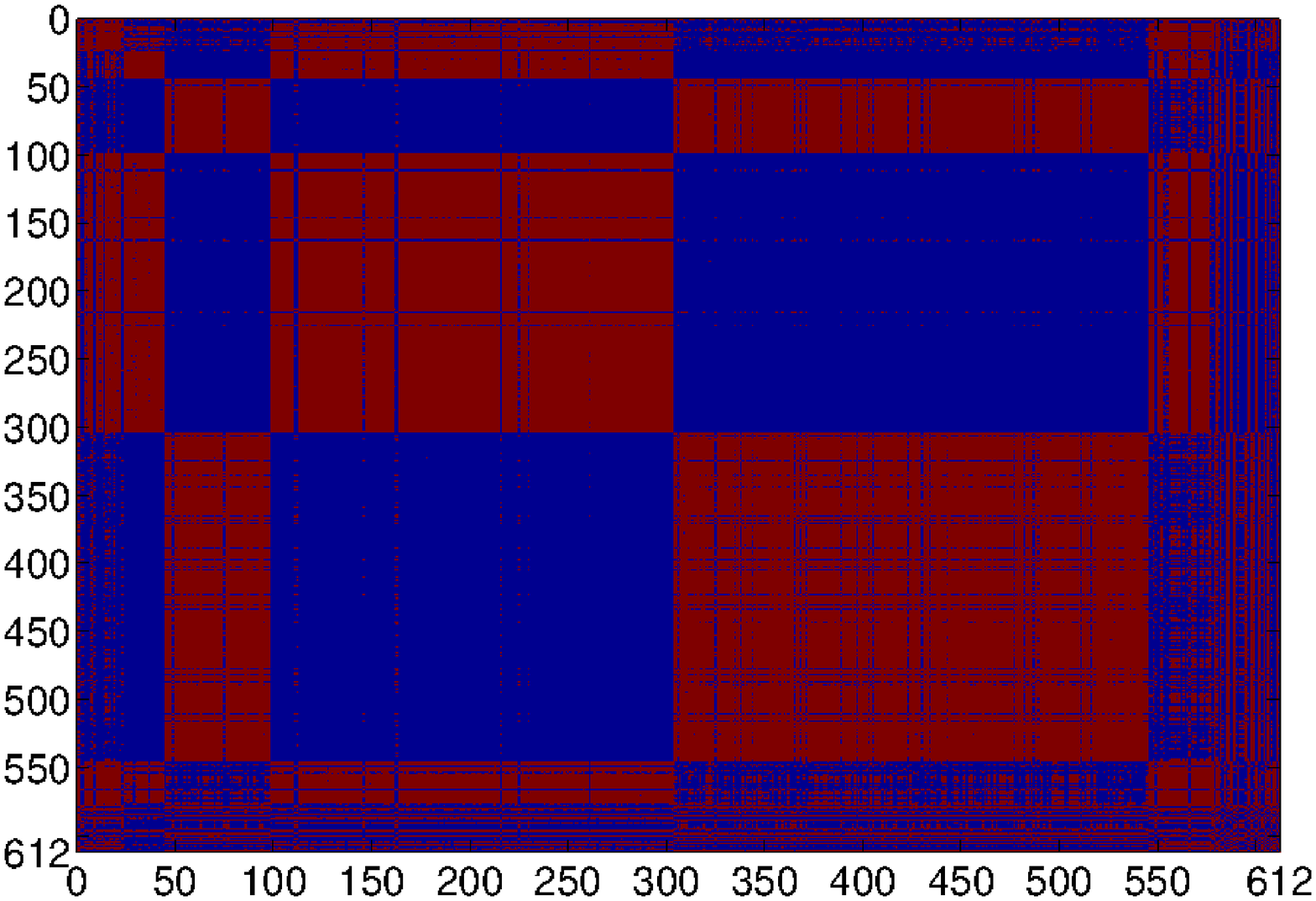}}

  \caption{Visualization of the binary similarity matrices sorted by party membership, for each of the six semesters. The row/column intervals corresponding to each party are the following: FL [1:23]; IDV [24:44]; LN [45:98]; PD [99:303]; PDL [304:521]; PT [522:545]; UDC [546:578]; Mixed [579:612].}
\label{matrici}
\end{figure}

\section{Network representation of Italian Parliament voting records}\label{net}

Given the similarity matrix $M$ among Parliamentarians, a network $\cal N$ can be built from $M$, by considering each Parliamentarian as a node of a weighted undirected graph $G=(V,E,\omega)$, where $V$ is the set of vertices, $E$ the set of edges,  and $\omega: E \mapsto \Re$  is a mapping that assigns a weight to the edge $(i,j)$, between the vertices $i$ and $j$. The weight $\omega(i,j)$ corresponds to  the similarity value $M_{ij}$ between Parliamentarians $i$ and $j$. $M$ can thus be considered the weighted adjacency matrix of $G$. Since weights of value zero are rather rare, the graph $G$ is an ``almost"  complete graph, and the application of network analysis methods could not uncover interesting properties. In order to study and investigate the Parliamentarian network, a thresholding operation on $M$ has been considered, i.e., fixed a threshold $\delta$,    let $B_{\delta}$ be the adjacency matrix of $G$ obtained by assigning a value equal to $1$ to a generic element $B_{ij}$ of $B$ if the corresponding value $M_{ij} \geq \delta$, $0$ otherwise. In the following, the subscript $\delta$ is omitted from the binary matrix $B$, when the value used for $\delta$ is clear from the context.

\subsection{Block visualization}
In order to visualize the similarity matrix \index{similarity matrix} $M$,  we considered the binarized matrix \index{binarized matrix} $B$ with $\delta= 0.6$.  $B$ has been then reordered such that Parliamentarians of the same party are located as consecutive rows/columns. 

Figure \ref{matrici}  shows how the two political parties PDL (rows 304:521) and LN (rows 45:98), that supported the center-right government, progressively reduce their intra-group similarity, while the opposition parties PD (rows 99:303), IDV (rows 24:44), and UDC (rows 546:578) present the opposite trend, i.e. in the first three semesters their intra-group similarity slightly diminishes, in the second three semesters, on the contrary, it increases. It is interesting to note that members of FL (rows 1:23) maintain their high similarity for all the periods, although they separated from PDL in 2010.   Another important observation regards the new formed group PT, whose Representatives come from the center-left parties. Although this was constituted in the sixth semester to avoid the government fall, its members showed a good political affinity since the first semester (rows/columns 522:545). The figures clearly show the boosting of agreement from the first to the last semester.

\section{Analysis of network structure}\label{struct}

The representation of similarity among Parliamentarians as a network $\cal N$ allows the analysis of topological features that characterize the network structure. In the following some measurements, coming from graph theory, that provide a quantitative characterization of the structural properties of the Parliamentarian networks \index{Parliamentarian networks},  are reported and discussed. 
 
As pointed out by Wasserman and Faust in  \cite{Wass2009}, a main goal in a network is the detection of the \emph{most important} or \emph{central} nodes. Measures introduced by researchers to interpret the concept of \emph{centrality} are based on the position of nodes in the network. Important nodes are usually located in strategic positions within the network. \emph{Degree} and \emph{betweenness} are two concepts, explained below,  that try to quantify the importance of nodes.   We first consider indices based on the degree concept.

\subsection{Density, degree centrality, and average degree}
A node that has many ties  has a central role since it can quickly exchange information with the other nodes of the network. The simplest index of centrality is the number of neighbors of a node, i.e. its degree. The \index{degree} degree $k_i$ of a vertex $i$ is defined as: 
\begin{equation}\label{}
k_i=\sum_jB_{ij}=\sum_jB_{ji}
\end{equation}
that is the number of edges connected to $i$. 

Other two measures of connectedness are \emph{average degree} and \emph{density} \index{average degree} \index{density}.  
The \emph{average degree} $\left \langle k \right \rangle$ is defined as:
\begin{equation}\label{}
\left \langle k \right \rangle=\frac{1}{\mid V \mid}\sum_i{k_i}=\frac{1}{\mid V \mid}\sum_{ij}B_{ij}
\end{equation}
i.e. it is the average of the degrees for all vertices in the network.  

The \emph{density} $d$ of a network is the number of links in the graph, expressed as a proportion of the maximum possible number of links. 
\begin{equation}\label{}
d = \frac{2 \mid E \mid }{\mid V \mid (\mid V \mid -1)}
\end{equation}
Network density depends on the size of the network. A more useful measure that allows to evaluate the structural cohesion of a network, independently of its size is the \emph{average degree}. 

\begin{figure}[ht]
  \centering
  \subfigure[\label{}]
  {\includegraphics[width=0.49\textwidth]{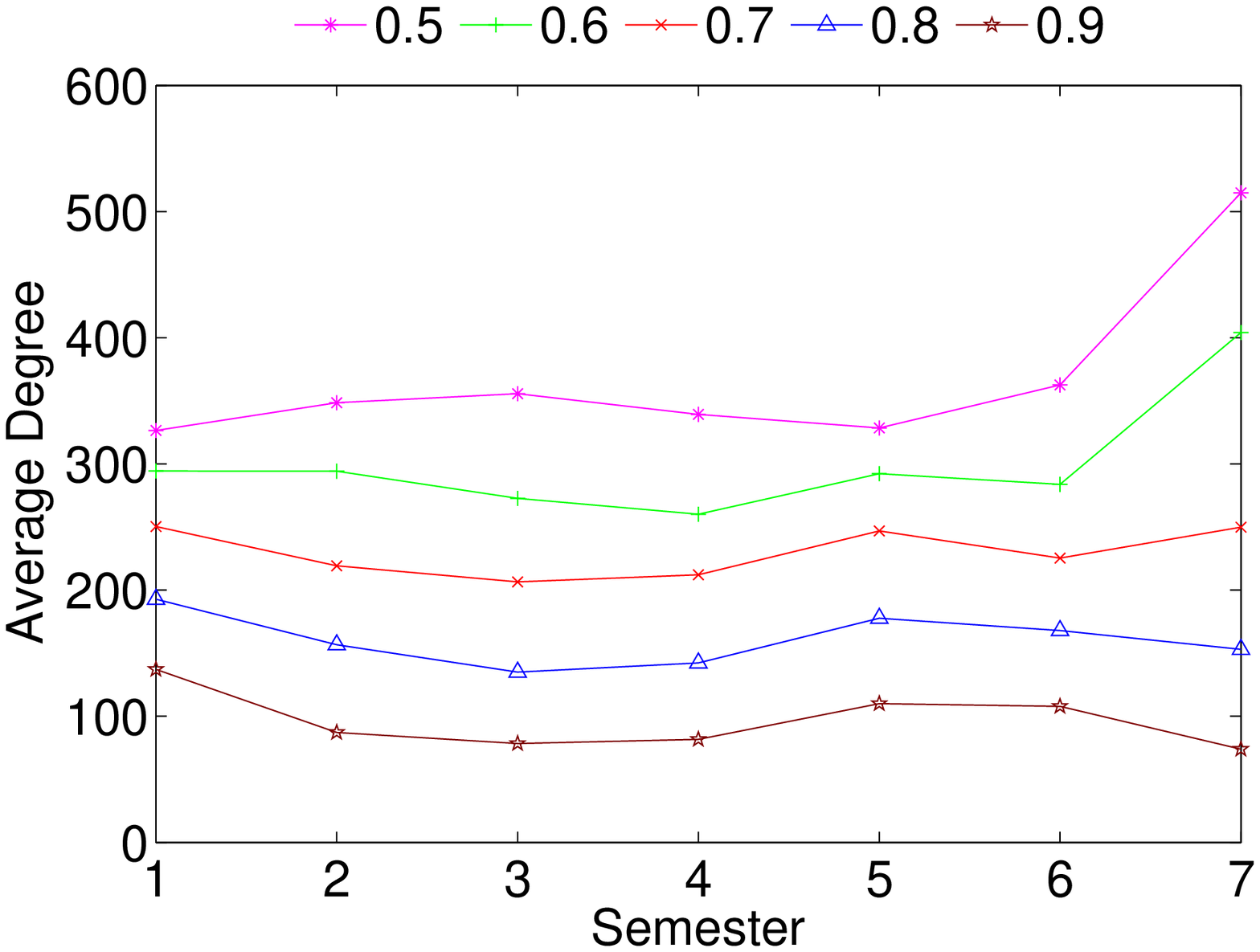}}
  \subfigure[\label{}]
  {\includegraphics[width=0.49\textwidth]{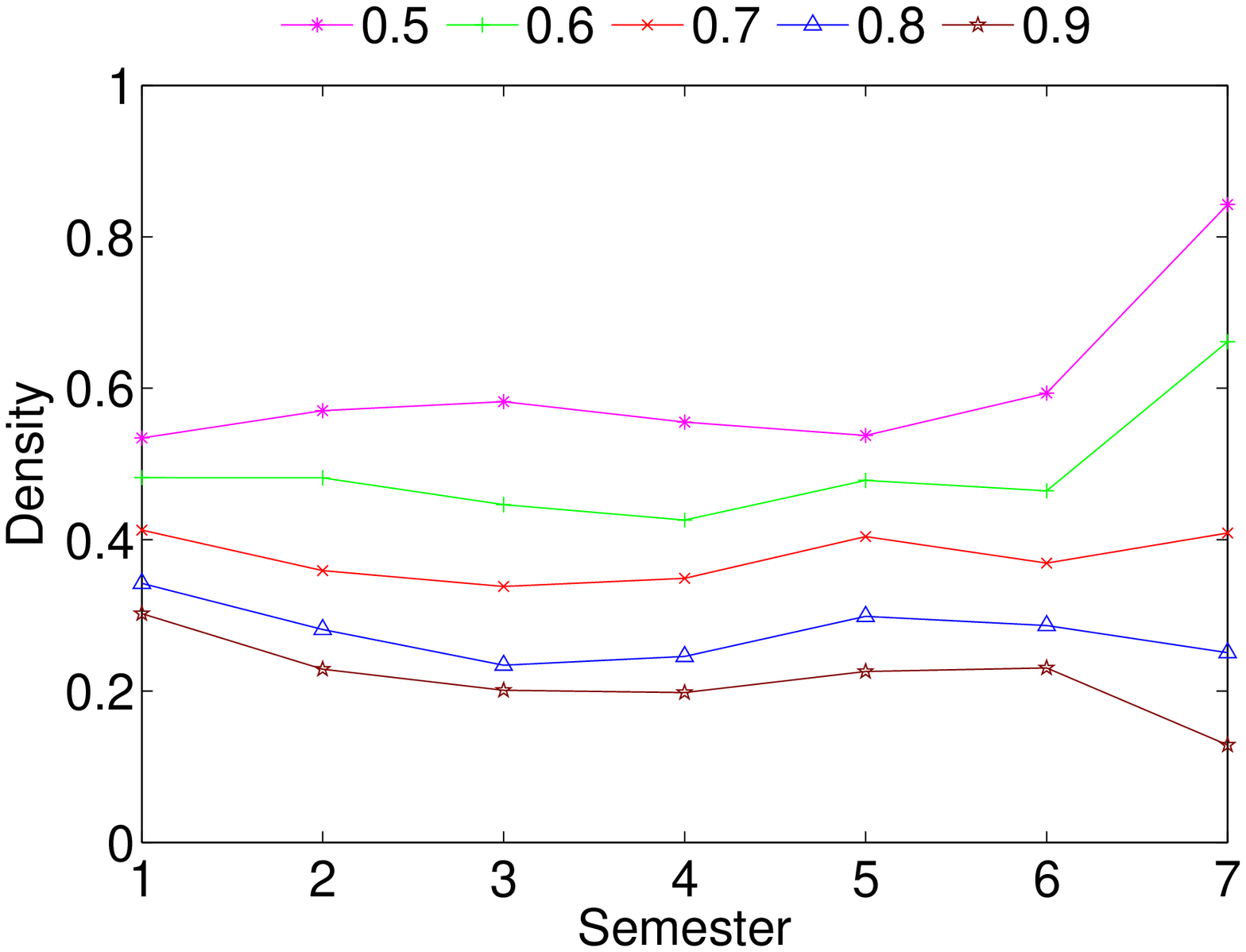}}

 \caption{Average degree (a) and density (b) on the Parliamentarian networks. Each measure is evaluated along the seven semesters by thresholding the corresponding similarity matrix $M$ for each semester at $0.5, 0.6, 0.7, 0.8, 0.9$.}
\label{dd}
\end{figure}

These last two concepts are strictly related. If a network has low density, the average connectivity of its vertices, i.e. $\left \langle k \right \rangle$, is low, thus there can be many isolated nodes and small connected components. As the number of edges increases, the connectivity increases too, until a unique component in which the vertices are connected to each other is present.

This behavior can be observed in Figure \ref{dd}, where density and average degree are computed along the seven semesters for different values of the threshold $\delta$.  The figure shows that lower values of $\delta$ imply a higher connectivity, which culminates in the seven semester for $\delta=0.5$, with an average degree above 500, and a density near 0.9. Considering that the number of nodes is 612, this means that Parliamentarians voted in a similar way in at least  50\% out of all the roll calls.

\begin{table}[t]
\caption{Parliamentarians with the highest degree centrality along the semesters, who are present in at least two out of seven semesters of the XVI Legislature.} \label{degreec}
\begin{center}
\scriptsize
\begin{tabular}{p{3cm}p{2cm}p{1cm}p{1cm}p{1cm}p{1cm}p{1cm}p{1cm}p{1cm}}
\hline\noalign{\smallskip}
Name&Political party&sem1 &sem2&sem3 &sem4 &sem5&sem6&sem7\\
\hline
Siegfried Brugger &Mixed group &362 &- &384&- &- &466&-\\
\hline
Stefano Stefani &LN &351 &- &359&341&345 &-&-\\
\hline
Maria Grazia Siliquini&PT &350 &- &359 &340 &-&-&-\\
\hline
Adolfo Urso &Mixed group &349 &371 &- &337 &-&-&-\\
\hline
Claudio Scajola&PDL &349 &370 &- &- &347&-&528\\
\hline
Carmelo Lo Monte&Mixed group &- &- &- &- &345&-&528\\
\hline
Karl Zeller&Mixed group &363 &- &369 &- &-&-&-\\
\hline
Silvio Berlusconi&PDL &353 &372 &- &- &-&-&-\\
\hline
Stefania  Craxi&PDL &351 &- &364 &- &-&-&-\\
\hline
Michela Brambilla &PDL &350 &371&- &-&-&-&-\\
\hline
Gianfranco  Miccich\'e &Mixed group &349 &373&- &-&-&-&-\\
\hline
Giulio Tremonti &PDL &348 &372&- &-&-&-&-\\
\hline
Giacomo Stucchi &LN &348 &-&- &-&348&-&-\\
\hline
Andrea Ronchi &Mixed group &348 &370&- &-&-&-&-\\
\hline
Stefania Prestigiacomo &PDL &348 &370&- &-&-&-&-\\
\hline
Guido Crosetto &PDL &- &367&- &-&347&-&-\\
\hline
Francesco Bosi &UDC &- &-&398&-&-&400&-\\
\hline
Gabriella Mondello &UDC &- &-&383 &-&-&385&-\\
\hline
Riccardo Migliori &PDL &- &-&362 &-&345&-&-\\
\hline
Edmondo Cirielli&PDL &- &-&359 &-&350&-&-\\
\hline
Luca Barbareschi &Mixed group&- &-&359 &338&-&-&-\\
\hline
Giorgio Jannone&PDL &- &-&358 &342&-&-&-\\
\hline
Giulia Cosenza &PDL &- &-&358 &-&346&-&-\\
\hline
Francesco Pionati &PT &- &-&- &344&-&-&528\\
\hline
\end{tabular}
\end{center}

\end{table}

 Degree centrality \index{Degree centrality} in the Parliamentarian network means that a Representative agreed with many others in voting bills, thus he/she shares political affinity with the neighbors and could influence their future voting. In Table \ref{degreec} the top 20 Parliamentarians having the highest degree centrality for at least two out of the seven semesters are reported. The index has been computed by fixing the threshold $\delta$=0.6. It is interesting to note that  Claudio Scaiola (who had the role of Minister in the government) is one of the most central person, for four out of seven semesters, in particular in the last semester, while the ex Prime minister Silvio Berlusconi was a central node only in the first two semesters.  
 
 \begin{figure}[ht]
  \centering
   \subfigure[\label{}]
  {\includegraphics[width=0.48\textwidth]{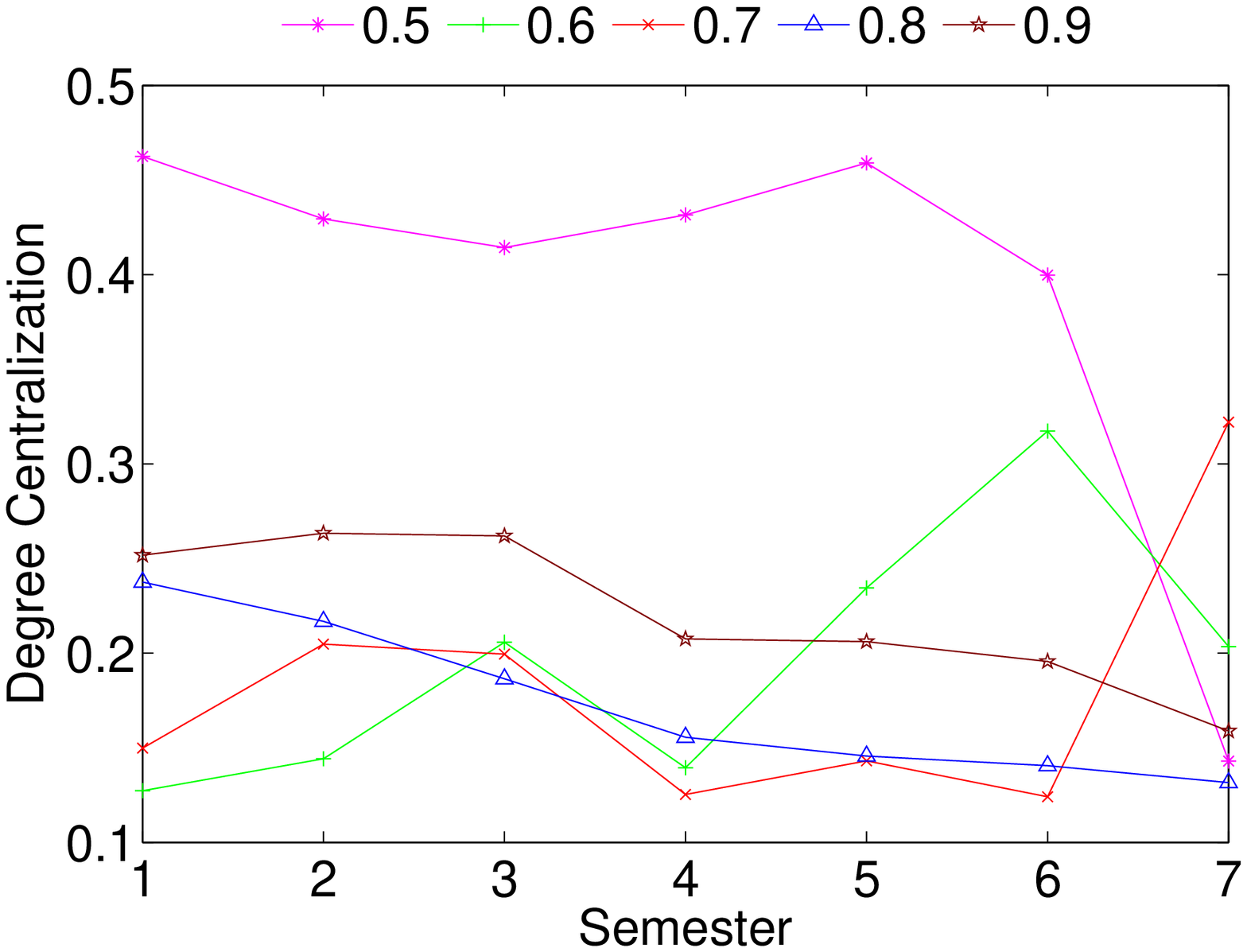}} 
  \subfigure[\label{}]
  {\includegraphics[width=0.48\textwidth]{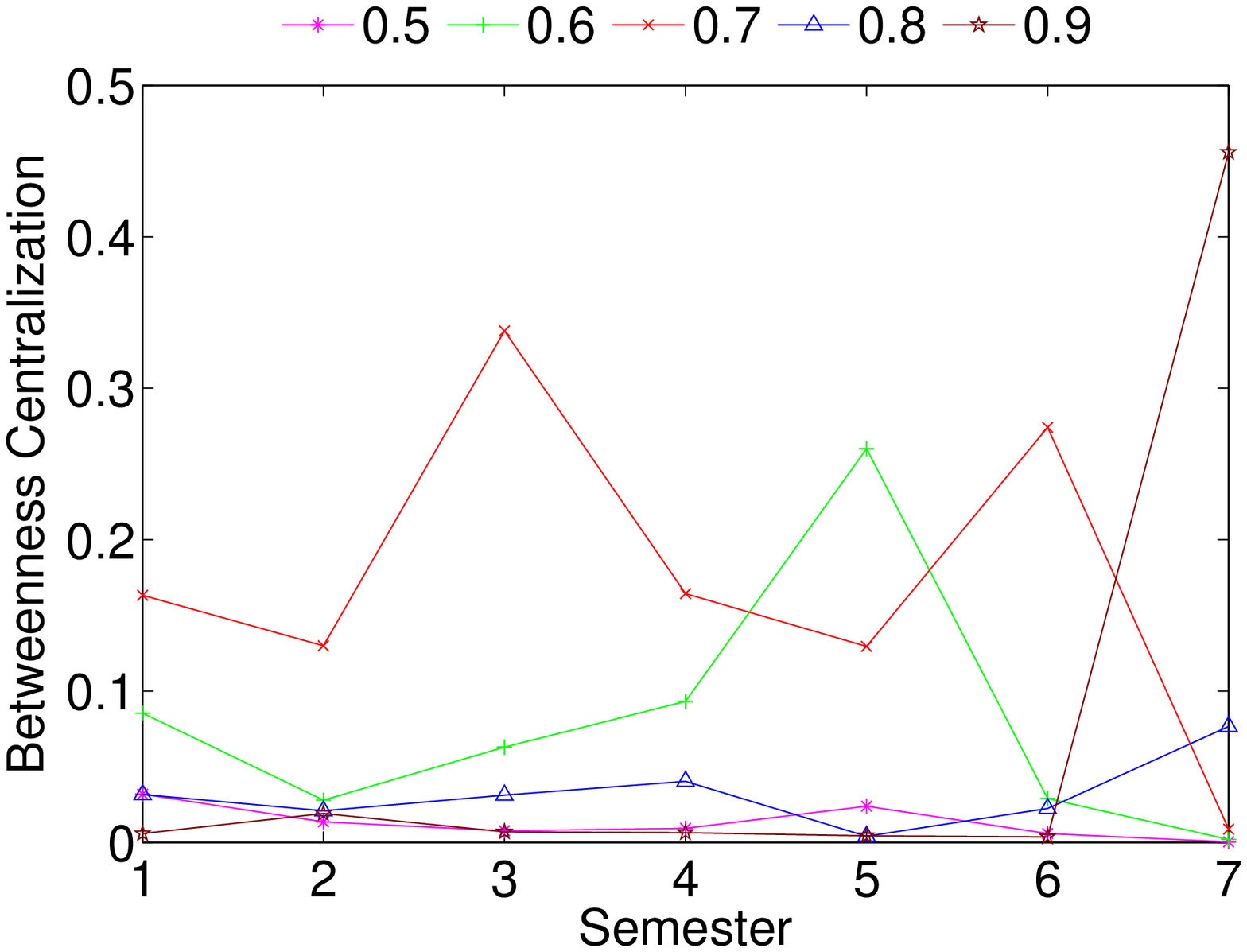}}

 \caption{Degree centralization (a) and betweenness centralization (b)  on the Parliamentarian networks. Each measure is evaluated along the seven semesters by thresholding the corresponding similarity matrix $M$ for each semester at $0.5, 0.6, 0.7, 0.8, 0.9$.}
\label{bdc}
\end{figure}

The set of degrees of all the nodes of a network can be summarized by a unique index named \emph{degree centralization} \index{degree centralization} that computes the variation in the degrees of vertices divided by the maximum degree variation which is possible in a network of the same size.
\begin{equation}\label{}
Cd({\cal N}) = \frac{1}{(\mid V \mid -1)(\mid V \mid -2)}\sum_i(Cd_{max} - Cd_i)
\end{equation}
where $Cd_{max}$ is the largest value of degree centrality in the network and $Cd_i$ is the degree centrality of vertex $i$, corresponding to the degree $k_i$.
Degree centralization is a measure of the dispersion of node degrees since it compares the degree of each node with the maximum degree present in the network.
Its value ranges from $0$, meaning that all degrees are equal, thus the graph has no variation,  to $1$, when a single node interacts with all the other $\mid V \mid -1$ nodes, while the other nodes are connected to only this one, which is the case of a star graph.
Figure \ref{bdc}(a) shows degree centralization along the seven semesters for increasing values of threshold $\delta$. We can note that the highest values are obtained with $\delta=0.5$, though in the seventh semester it drastically drops, analogously to the values computed for the other thresholds. However, most of the degree centralization values range between 0.1 and 0.26, except for $\delta=0.6$ in the sixth semester, and $\delta=0.7$ in the seventh semester, indicating that the network is rather regular, i.e. the degrees of all nodes are similar. 

\begin{table}[h!]
\caption{The 20 Parliamentarians with the highest betweenness centrality in the network of the 3rd semester with $\delta=0.7$.} \label{sem307}

\begin{center}
\scriptsize
\begin{tabular}{p{3cm}p{2cm}p{1.5cm}}
\hline\noalign{\smallskip}
Name&Political party&Betweenness centrality\\
\hline
Gabriella Mondello&UDC&0.3394\\
\hline
Francesco Laratta&PD&0.0322\\
\hline
Erminio Angelo Quartani&PD&0.0286\\
\hline
Alessandro Naccarato&PD&0.0269\\
\hline
Dario Franceschini&PD&0.0261\\
\hline
Armando Dionisi&UDC&0.0255\\
\hline
Gian Luca Galletti&UDC&0.0242\\
\hline
Angelo Compagnon&UDC&0.0240\\
\hline
Antonello Giacomelli&PD&0.0233\\
\hline
Nedo Lorenzo Poli&UDC&0.0229\\
\hline
Lorenzo Ria&UDC&0.0219\\
\hline
Michele Pompeo Meta&PD&0.0195\\
\hline
Roberto Rao&UDC&0.0189\\
\hline
Roberto Occhiuto&UDC&0.0177\\
\hline
Giuseppe Ruvolo&PT&0.0165\\
\hline
Antonio De Poli&UDC&0.0158\\
\hline
Nunzio Francesco Testa&UDC&0.0152\\
\hline
Mario Tassone&UDC&0.0144\\
\hline
Italo Tanoni&Mixed group&0.0142\\
\hline
Anna Teresa Formisano&UDC&0.0129\\
\hline
\end{tabular}
\end{center}
\end{table}

\begin{table}[h!]
\caption{The 20 Parliamentarians with the highest betweenness centrality in the network of the 6th semester with $\delta=0.7$.} \label{sem607}
\begin{center}
\scriptsize
\begin{tabular}{p{3cm}p{2cm}p{1.5cm}}
\hline\noalign{\smallskip}
Name&Political party&Betweenness centrality\\
\hline
Carmine Santo Patarino&FL&0.2762\\
\hline
Roberto Rosso&PDL&0.2758\\
\hline
Luca Volont\'e&UDC&0.1153\\
\hline
Mario Baccini&PDL&0.1060\\
\hline
Francesco Divella&FL&0.0243\\
\hline
Marco Fedi&PD&0.0182\\
\hline
Maurizio Migliavacca&PD&0.0180\\
\hline
Federica Mogherini Rebesani&PD&0.0172\\
\hline
Angelo Compagnon&UDC&0.0117\\
\hline
Gabriella Mondello&UDC&0.0114\\
\hline
Anna Teresa Formisano&UDC&0.0114\\
\hline
Gianfranco Paglia&FL&0.0112\\
\hline
Angela Napoli&FL&0.0112\\
\hline
Adolfo Urso&Mixed group&0.0109\\
\hline
Benedetto Della Vedova&FL&0.0108\\
\hline
Enzo Carra&UDC&0.0108\\
\hline
Lorenzo Ria&UDC&0.0105\\
\hline
Roberto Rao&UDC&0.0102\\
\hline
Angelo Cera&UDC&0.0101\\
\hline
Pierluigi Mantini&UDC&0.0099\\
\hline
\end{tabular}
\end{center}
\end{table}

\begin{table}[h!]
\caption{The 20 Parliamentarians with the highest betweenness centrality in the network of the 7th semester with $\delta=0.9$.} \label{sem709}
\begin{center}
\scriptsize
\begin{tabular}{p{3cm}p{2cm}p{1.5cm}}
\hline\noalign{\smallskip}
Name&Political party&Betweenness centrality\\
\hline

Fiamma Nirenstein&PDL&0.4639\\
\hline
Gaetano Porcino&IDV&0.4587\\
\hline
Matteo Mecacci&PD&0.4583\\
\hline
Massimo Parisi&PDL&0.4568\\
\hline
Nicodemo Nazzareno Oliverio&PD&0.4021\\
\hline
Arturo Iannaccone&PT&0.1973\\
\hline
Vincenzo Barba&PDL&0.1915\\
\hline
Luigi Vitali&PDL&0.1846\\
\hline
Paolo Bonaiuti&PDL&0.1322\\
\hline
Francesco Colucci&PDL&0.0811\\
\hline
Michela Vittoria Brambilla&PDL&0.0510\\
\hline
Maurizio Bernardo&PDL&0.0496\\
\hline
Riccardo De Corato&PDL&0.0445\\
\hline
Gregorio Fontana&PDL&0.0384\\
\hline
Valentino Valentini&PDL&0.0364\\
\hline
Sestino Giacomoni&PDL&0.0364\\
\hline
Giampaolo Fogliardi&PD&0.0355\\
\hline
Giorgio Merlo&PD&0.0352\\
\hline
Agostino Ghiglia&PDL&0.0306\\
\hline
Ida D'Ippolito Vitale&PDL&0.0306\\
\hline
\end{tabular}
\end{center}
\end{table}

\subsection{Betweenness}

If two nodes are not directly connected with an edge, their possibility of interacting depends on the paths between them, thus on the nodes constituting these paths. A node, then, can be considered central if it appears in the shortest paths joining many of the other nodes. The \emph{betweenness centrality} \index{betweenness centrality} is defined as:

\begin{equation}\label{}
B_i = \sum_{j,k, j \neq k} \frac{n_{jk}(i)}{n_{jk}}
\end{equation}
where $n_{jk}(i)$ is the number of shortest paths between vertices $j$ and $k$ that pass through vertex $i$, and $n_{jk}$ is the total number of shortest paths between $j$ and $k$.
The \emph{betweenness centrality} of a node measures the importance of a vertex  in the network in terms of number of shortest paths in which that vertex  participates.

Analogously to degree centralization, \emph{betweenness centralization} \index{betweenness centralization} measures the betweenness centrality variation  with respect to the maximum possible variation in node betweenness, and it is defined as:

\begin{equation}\label{}
B_{\cal N} = \frac{1}{\mid V \mid -1}\sum_i(B'_{max} - B'_i)
\end{equation}
where $B'_i$ is  $B_i$ normalized with respect to the maximum reachable value, and $B'_{max}$ is the largest value of betweenness centrality in the network, standardized as $B_i$. The index reaches its maximum value, equals to $1$, if the network is a star graph. In fact, in the star graph, the vertex in the middle has the highest betweenness centrality because it is on every geodesic, while all the other vertices have betweenness centrality of 0 as they are on no geodesics. On the other hand, the minimum value of $B_{\cal N}$, which is $0$, occurs when all the vertices have the same betweenness centrality. 
Figure \ref{bdc}(b) reports the betweenness network centrality along the seven semesters for increasing values of $\delta$. The figure points out that betweenness values are rather low except at the fifth semester for $\delta=0.6$, the third and sixth semesters for $\delta=0.7$, and seventh semester for $\delta=0.9$.  
Tables \ref{sem307}, \ref{sem607}, \ref{sem709} report the Parliamentarians having the top 20 highest values of betweenness for the third and sixth semester with $\delta=0.7$, and  seventh semester with $\delta=0.9$. These people should be the most influential Parliamentarians in the network and their removal could reduce communication among the groups.

\subsection{Clustering coefficient}

The clustering coefficient, \index{clustering coefficient} also known as transitivity, expresses the idea that two friends with a common friend are likely to be friends. This concept has been defined
by Watts and Strogatz in \cite{watt98}, and, in terms of network topology, it measures the number of triangles, i.e. the set of three vertices connected to each other. 
Given a node $i$, let $nt_i$ be the
number of links connecting the $k_i$ neighbors of $i$ to each
other.  The clustering coefficient of a node $i$ is defined as:
\begin{equation}\label{}
CC_i = \frac{2nt_i}{k_i(k_i -1)}
\end{equation} 
$nt_i$ represents the number of triangles passing through
$i$, and $k_i(k_i -1)/2$ the number of possible triangles that
could pass through node $i$. The clustering coefficient $CC$ of a graph is
the average of the clustering coefficients of the nodes it
contains:
\begin{equation}\label{}
CC=\frac{1}{\mid V \mid}{\sum_i CC_i}
\end{equation}

\begin{figure}[t]
\includegraphics[scale=.30]{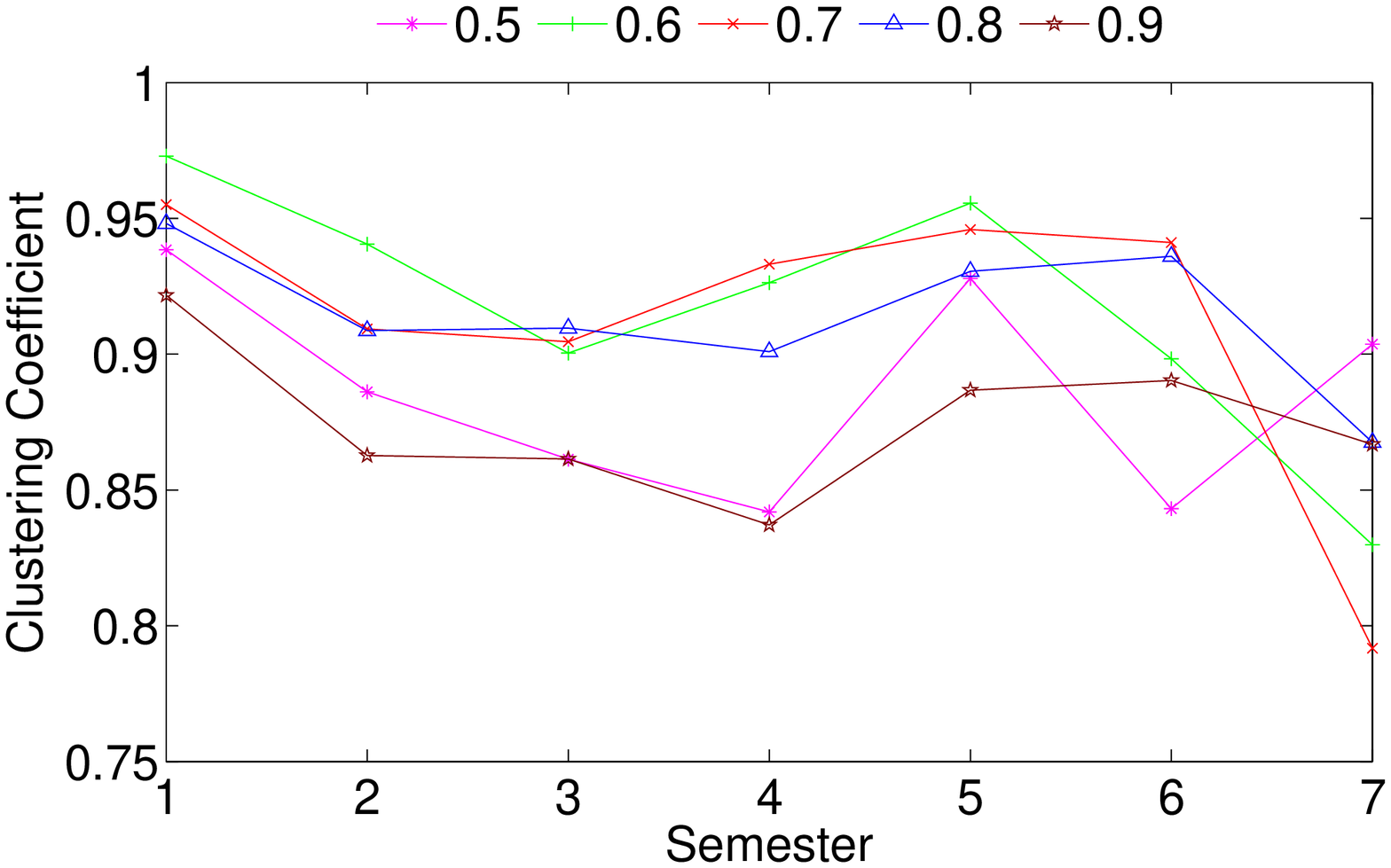}
\caption{Clustering coefficient of the Parliamentarian networks, evaluated along the seven semesters by thresholding the corresponding similarity matrix $M$ for each semester at $0.5, 0.6, 0.7, 0.8, 0.9$.}
\label{cc}      
\end{figure}

Clustering coefficient varies between $0$ and $1$. Figure \ref{cc} points out that the clustering coefficient is rather high, independently of the threshold $\delta$ used, showing thus that there are many triples of Parliamentarians voting in a similar manner.  However, it worth to note that there is no monotonicity between increasing the threshold $\delta$ and clustering coefficient values. The explanation of this behavior comes from the decreasing  degree values of vertices when $\delta$ augments. In fact, when $\delta=0.5$, average node degree is between 330 and 520, thus the number of possible triangles that could pass through a vertex is very high. This could drop the value of $CC_i$ if the neighbors of node $i$ are not well connected. On the other hand,  for $\delta=0.9$ both average degree and number of connections are rather low, thus clustering coefficient assumes smaller values.

 \begin{figure}[ht!]
  
  \subfigure[I Semester\label{}]
  {\includegraphics[width=0.7\textwidth]{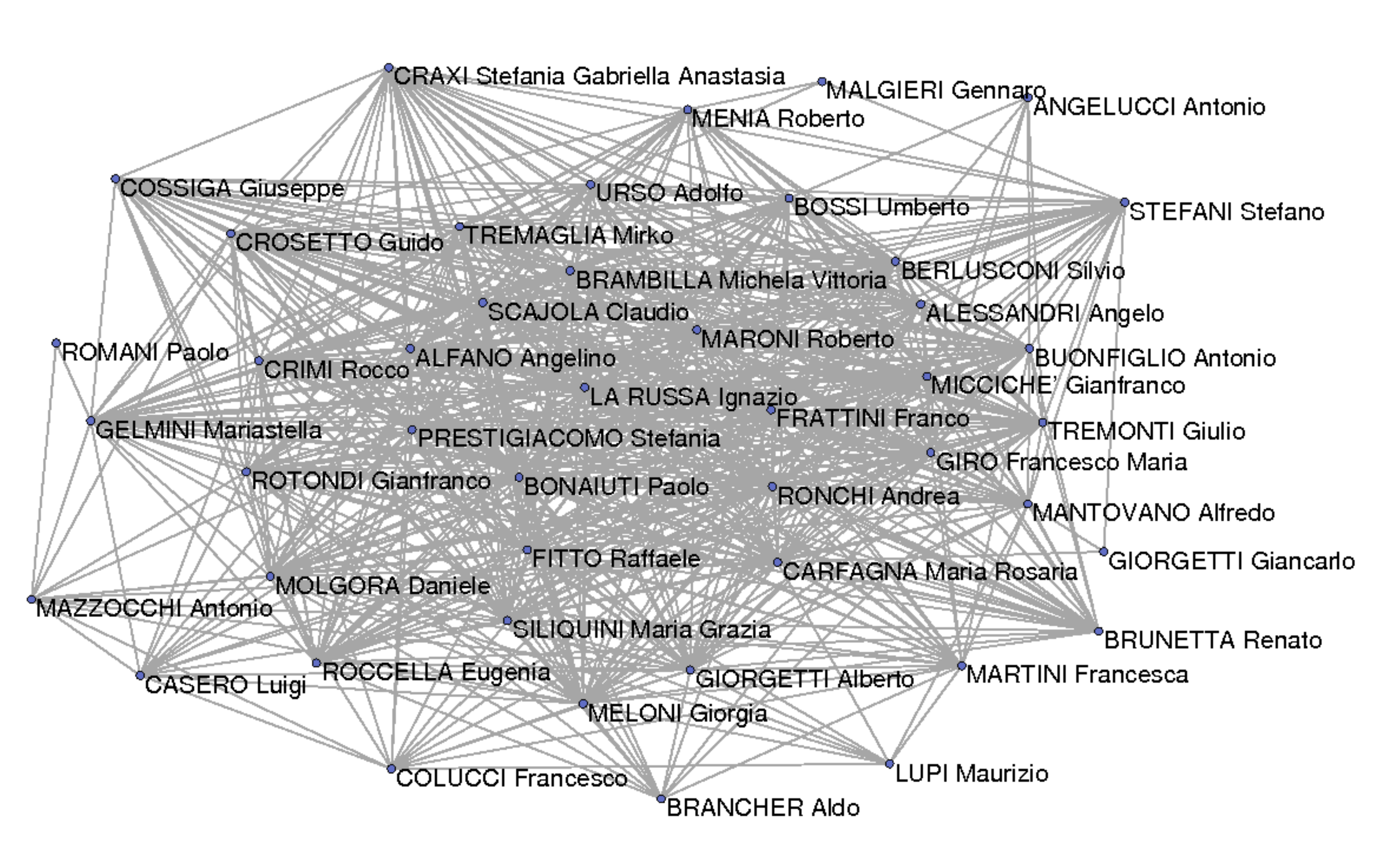}}\\
  \subfigure[II Semester \label{}]
  {\includegraphics[width=0.7\textwidth]{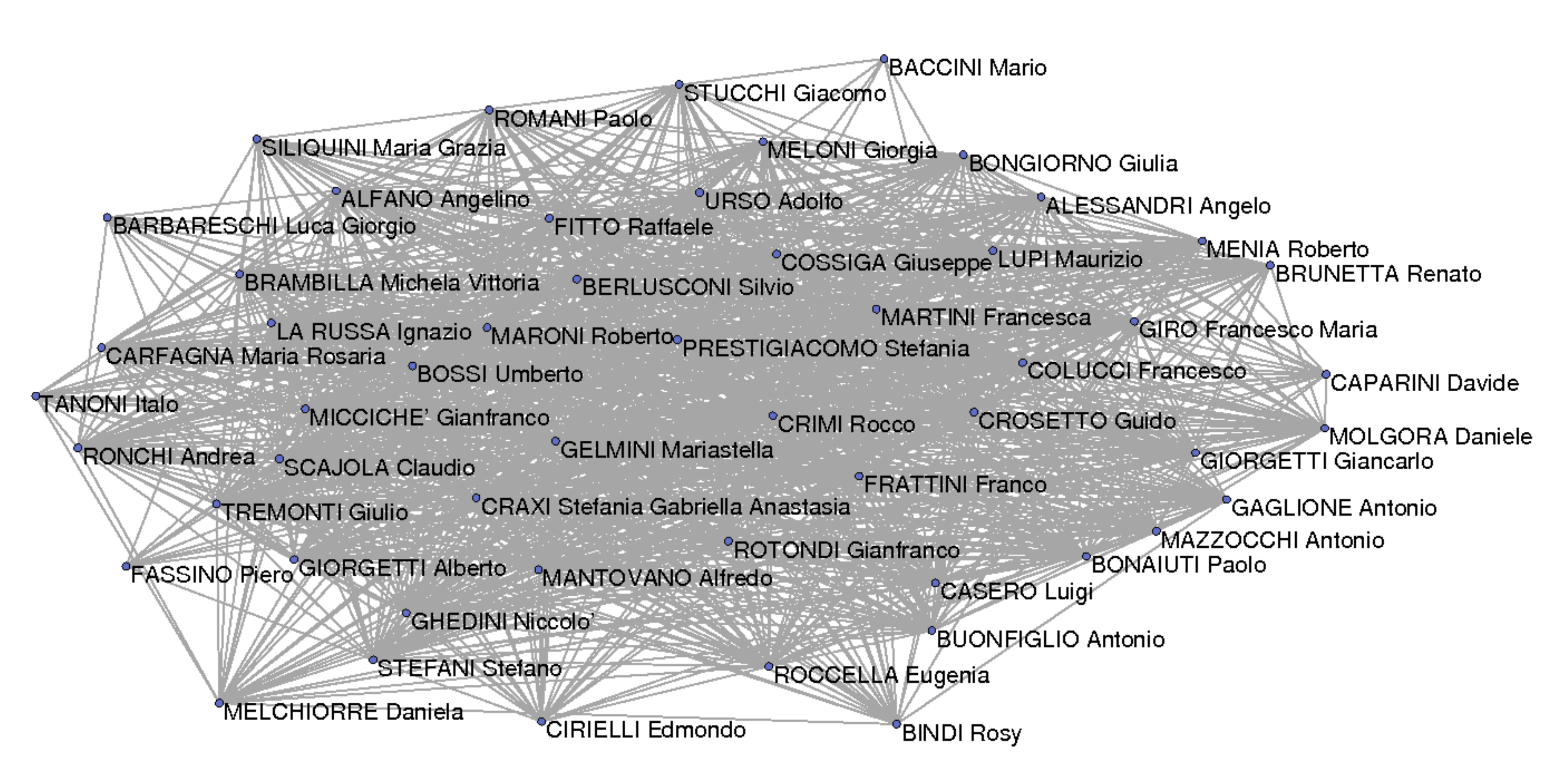}}\\
   \subfigure[III Semester \label{}]
  {\includegraphics[width=0.7\textwidth]{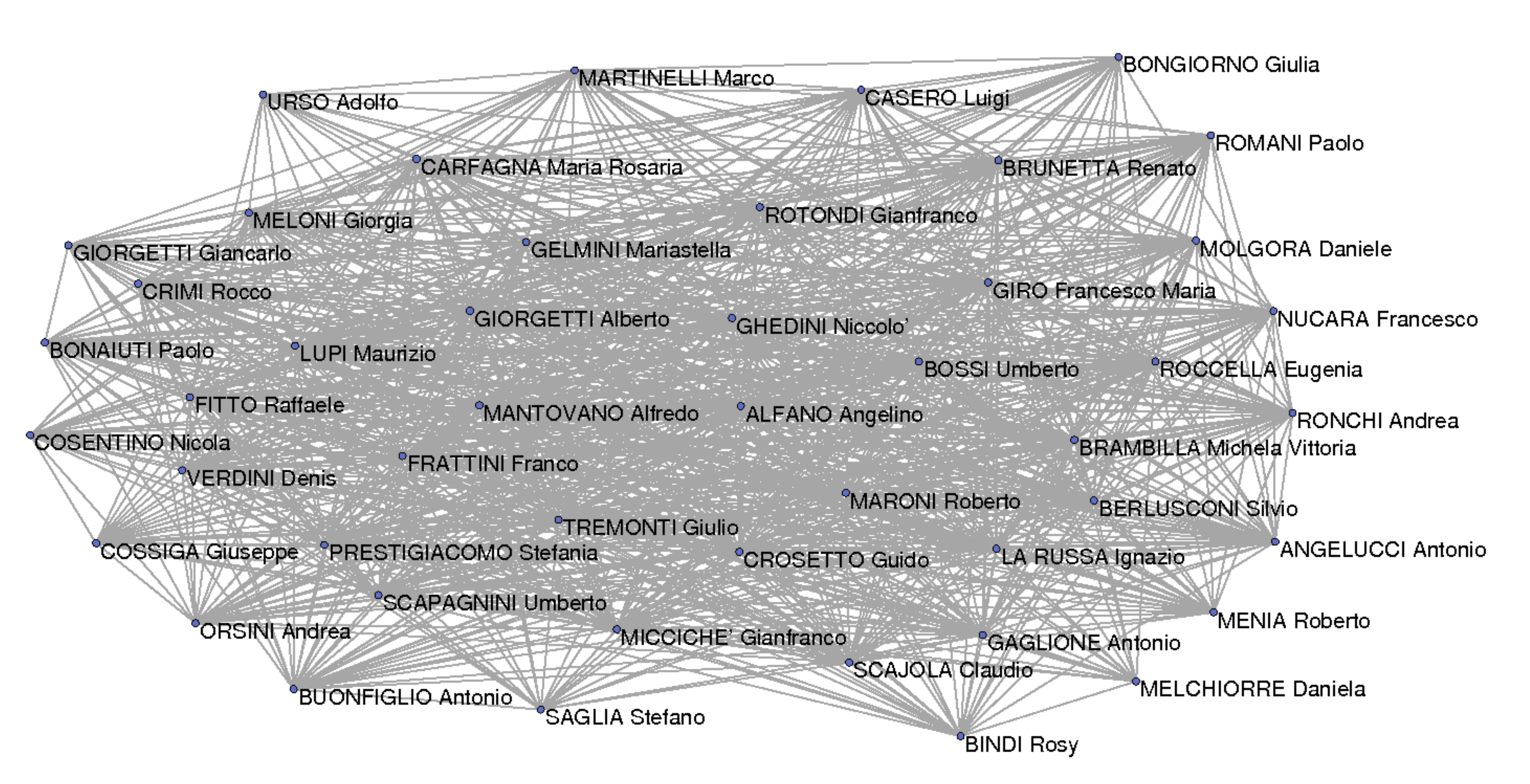}}
  \caption{p-cliques obtained for the I, II, and III semesters with number of Parliamentarians 45, 53, and 47 respectively. }
\label{clique123}
\end{figure}

 \begin{figure}[ht!]
 
  \subfigure[IV Semester\label{}]
  {\includegraphics[width=0.7\textwidth]{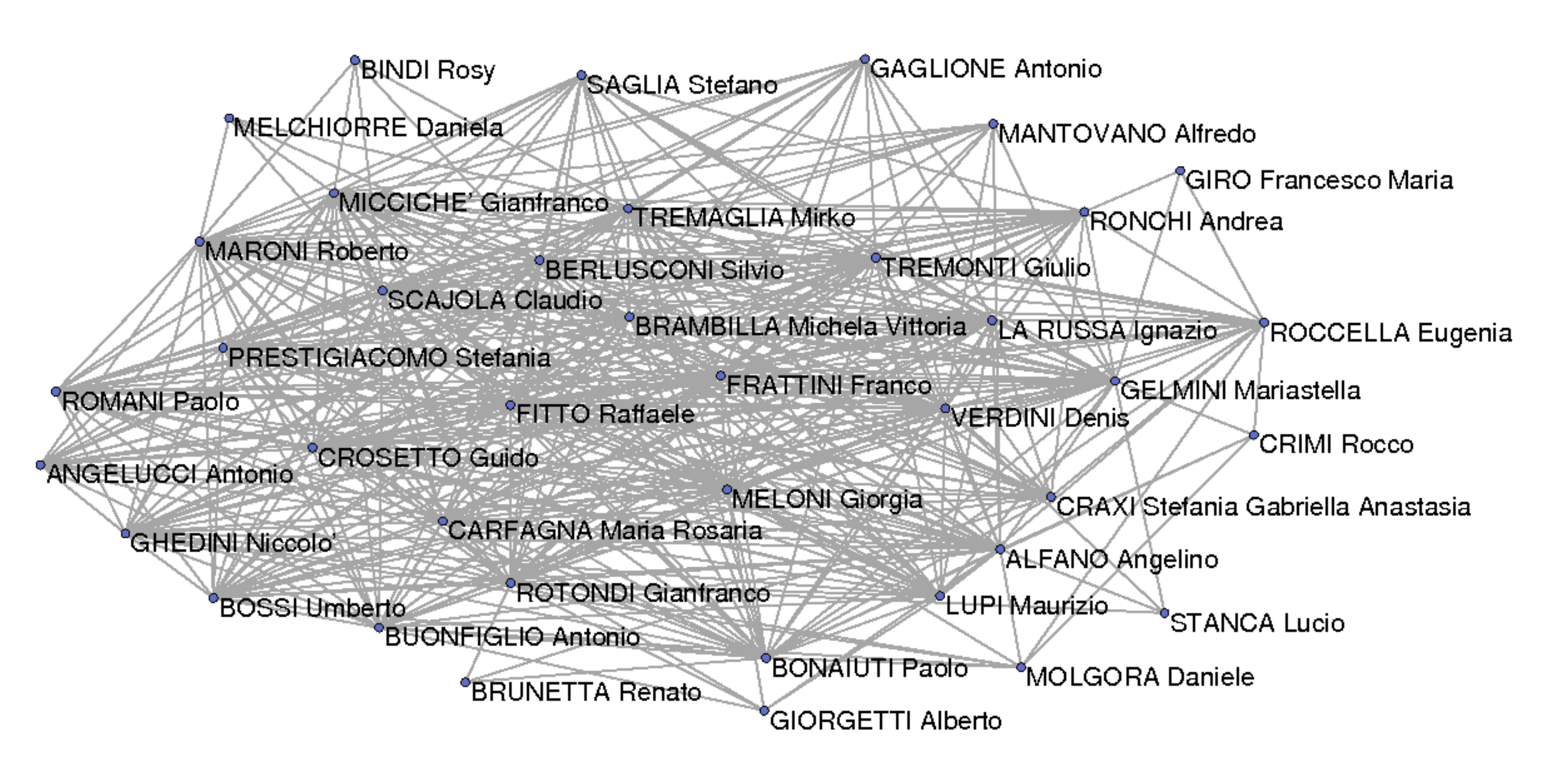}}\\
\subfigure[V Semester \label{}]
  {\includegraphics[width=0.7\textwidth]{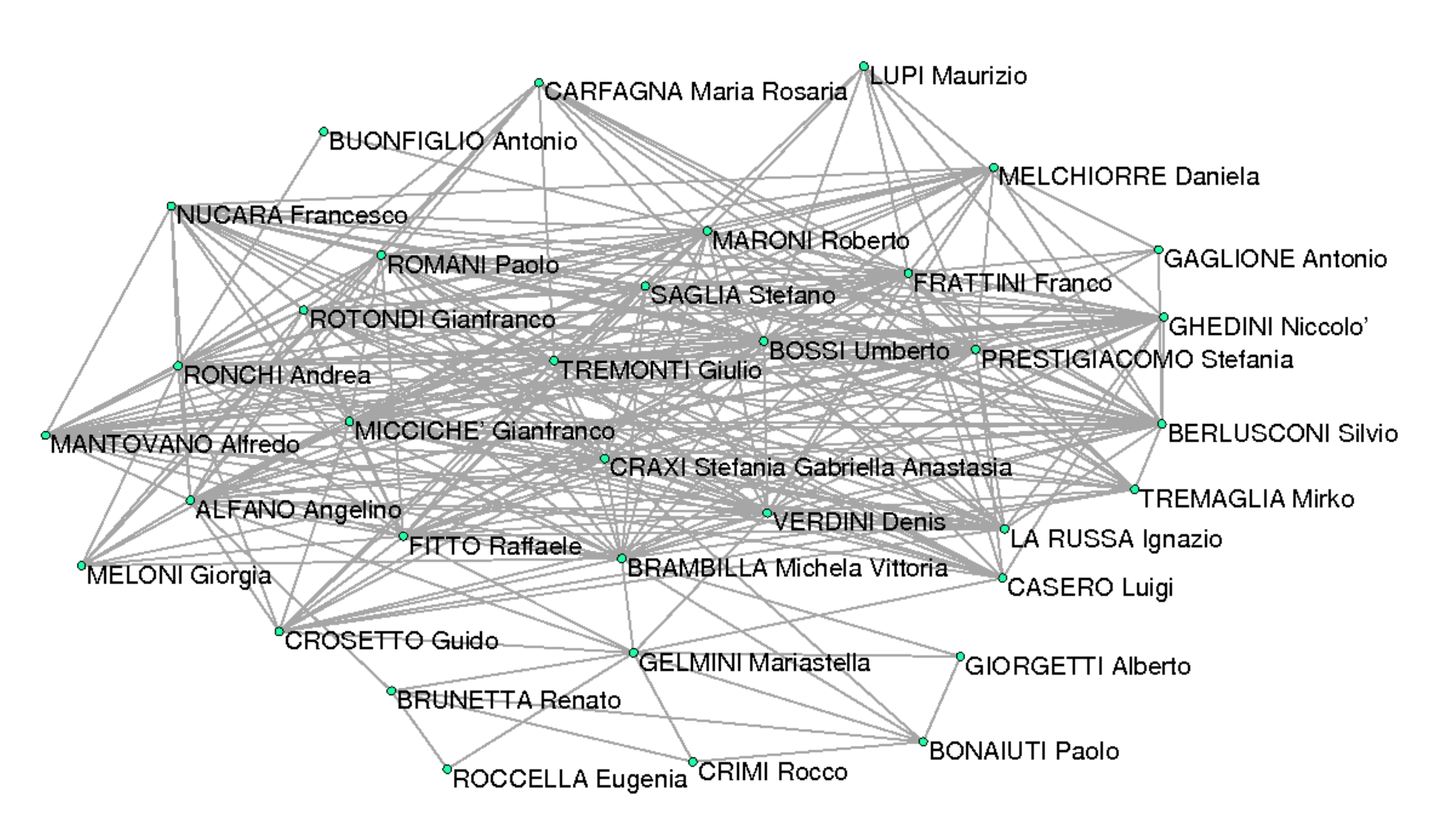}}\\
  \subfigure[VI Semester \label{}]
  {\includegraphics[width=0.7\textwidth]{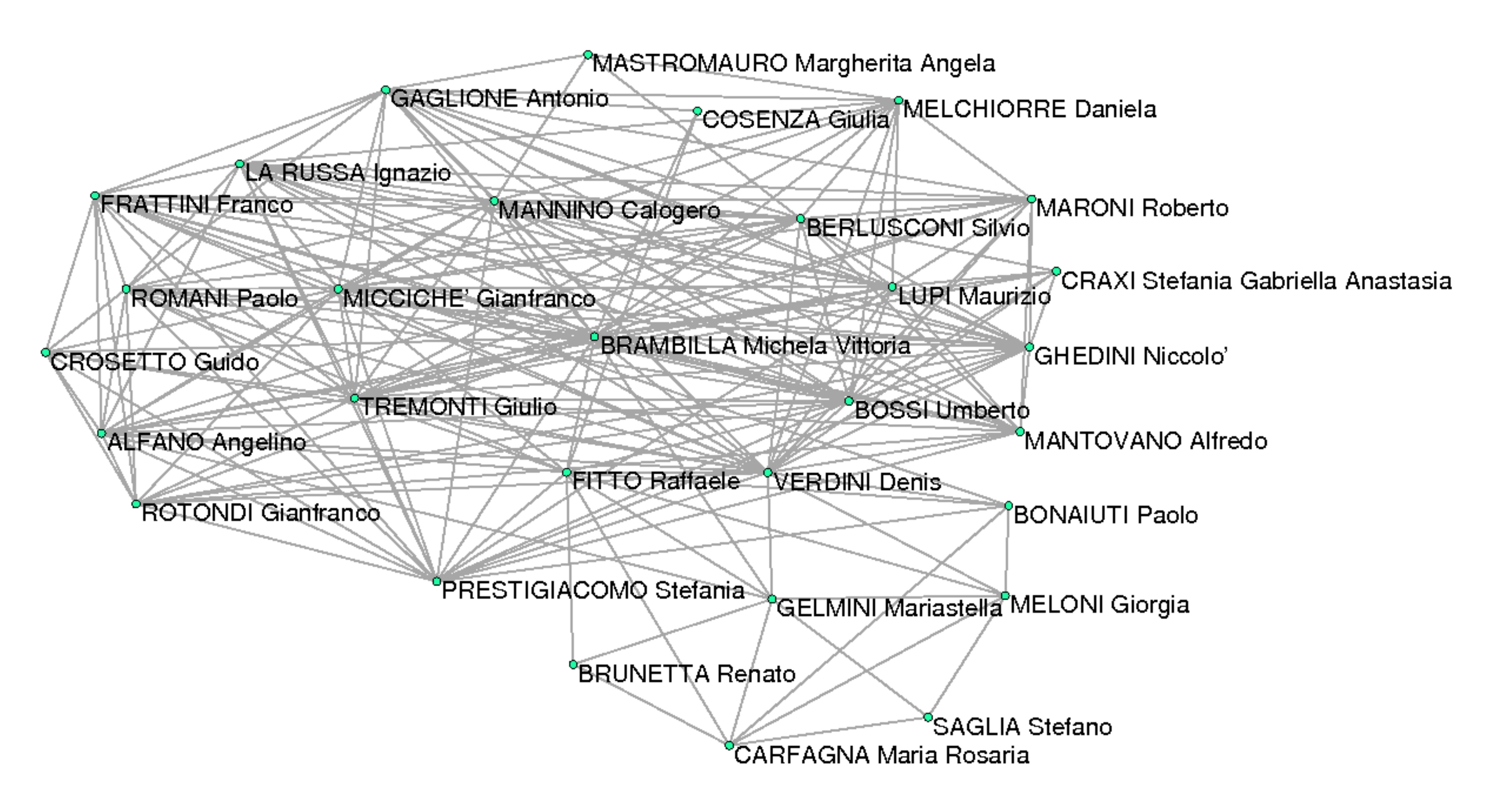}}
   \caption{p-cliques obtained for the IV, V, and VI semesters with number of Parliamentarians 39, 35, and 30 respectively. }
\label{clique456}
\end{figure}

 \begin{figure}[ht!]
  
   \subfigure[VII Semester \label{}]
  {\includegraphics[width=0.7\textwidth]{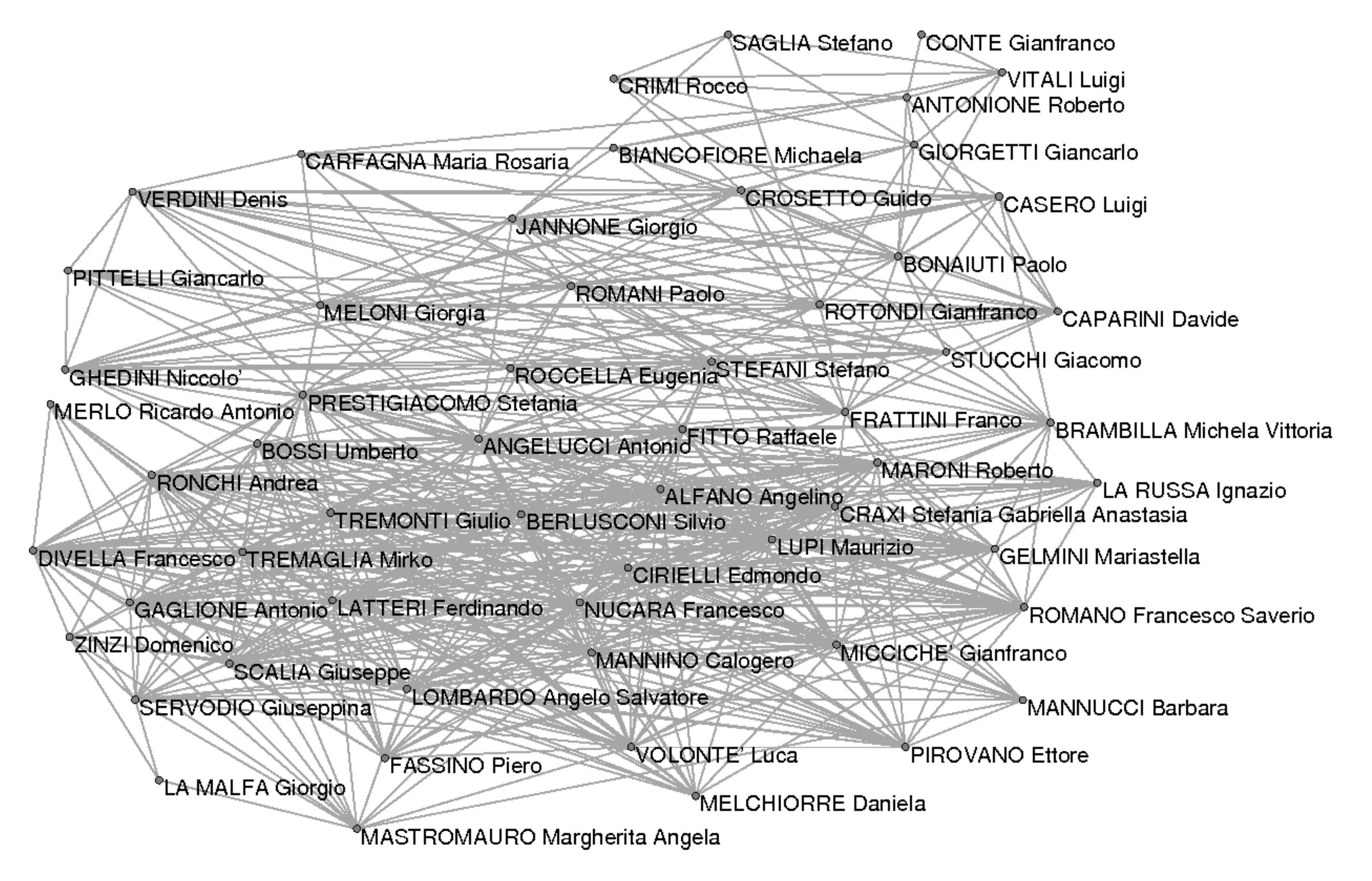}}
 \caption{p-cliques obtained for the VII semester composed by 58 Parliamentarians. }
\label{clique7}
\end{figure}

\subsection{p-cliques}
 
 Another important feature to study in networks is the presence of $cliques$, \index{cliques} i.e. maximal complete subgraphs of at least three nodes. However,  the request that each node must be connected with all the other nodes of the subgraph is rather strong. Thus we fix the attention on the \emph{p-cliques}, i.e. groups of nodes having  at least a proportion $p$ of neighbors inside the same group.  We considered PDL party, and computed the \emph{p-cliques}, with $p=0.5$. Very interestingly, we obtained a group of 19 Parliamentarians, reported in Table \ref{fedeli},  that remained compact for all the seven semesters. These Representatives have main roles in the PDL party. In particular, Angelino Alfano is the party secretary, Umberto Bossi  was the LN party secretary, while Roberto Maroni is the actual LN secretary and Minister of the Berlusconi's government, 9 out of 19 played the role of  Minister,  and the others have a main position in PDL party. The nucleus of 19 persons can be considered the ``most faithful" supporters of the ex-Prime Minister, that remained ``devoted" until the end. A variable number of other Representatives joined or left this dense group of people.   Figures \ref{clique123}, \ref{clique456}, and \ref{clique7} show the ``birth" and evolution of these p-cliques within the seven semesters. It is worth to note that in the fifth and sixth semester many Parliamentarians disappeared from this group of faithful supporters, while in the seventh the p-clique again increased with new entries.

\begin{table}[t]
\caption{The 19  most faithful Parliamentarians of ex-Prime Minister.} \label{fedeli}
\begin{center}
\scriptsize

\begin{tabular}{p{3cm}p{2cm}}
\hline\noalign{\smallskip}
Name&Political party\\
\hline
ALFANO Angelino & PDL\\
\hline
BERLUSCONI Silvio& PDL\\
\hline
BONAIUTI Paolo& PDL\\
\hline
BOSSI Umberto& LN\\
\hline
BRAMBILLA Michela& PDL\\
\hline
CARFAGNA Maria Rosaria& PDL\\
\hline
CROSETTO Guido& PDL\\
\hline
FITTO Raffaele& PDL\\
\hline
FRATTINI Franco& PDL\\
\hline
GELMINI Mariastella& PDL\\
\hline
LA RUSSA Ignazio& PDL\\
\hline
LUPI Maurizio& PDL\\
\hline
MARONI Roberto& LN\\
\hline
MELONI Giorgia& PDL\\
\hline
MICCICHE' Gianfranco& PDL\\
\hline
PRESTIGIACOMO Stefania& PDL\\
\hline
ROMANI Paolo& PDL\\
\hline
ROTONDI Gianfranco& PDL\\
\hline
TREMONTI Giulio& PDL\\
\hline
\end{tabular}
\end{center}

\end{table}

\begin{figure}[b]
\includegraphics[scale=.50]{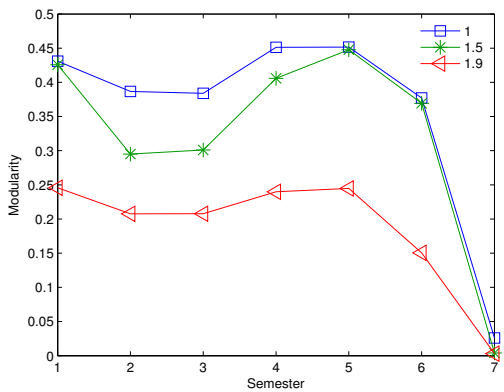}
\caption{Modularity for all semesters with different values of $\gamma$.}
\label{mod}    
\end{figure}

\begin{figure}[ht!]
  \centering
  \subfigure[I Semester]
  {\includegraphics[width=0.5\textwidth]{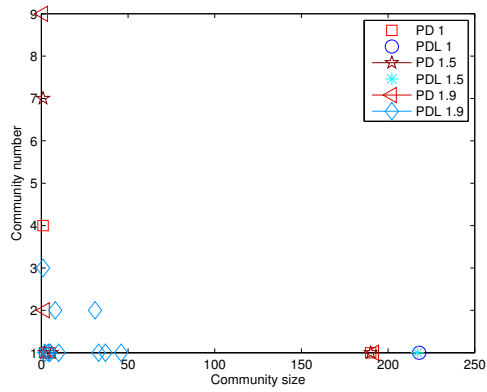}}
  \subfigure[II Semester\label{}]
  {\includegraphics[width=0.5\textwidth]{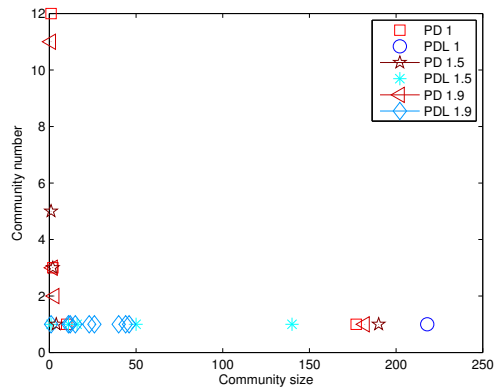}}
  \subfigure[III Semester\label{}]
  {\includegraphics[width=0.5\textwidth]{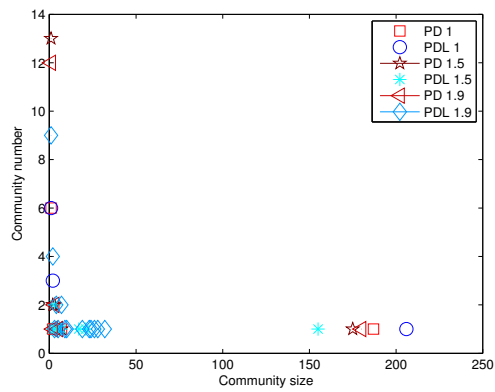}}
   \caption{Number of communities in which the two main parties PDL and PD are split and respective size for the first three semesters, for different values of $\gamma$.}
\label{comm1}
\end{figure}
  
  \begin{figure}[ht!]
  \centering
  \subfigure[IV Semester\label{}]
  {\includegraphics[width=0.5\textwidth]{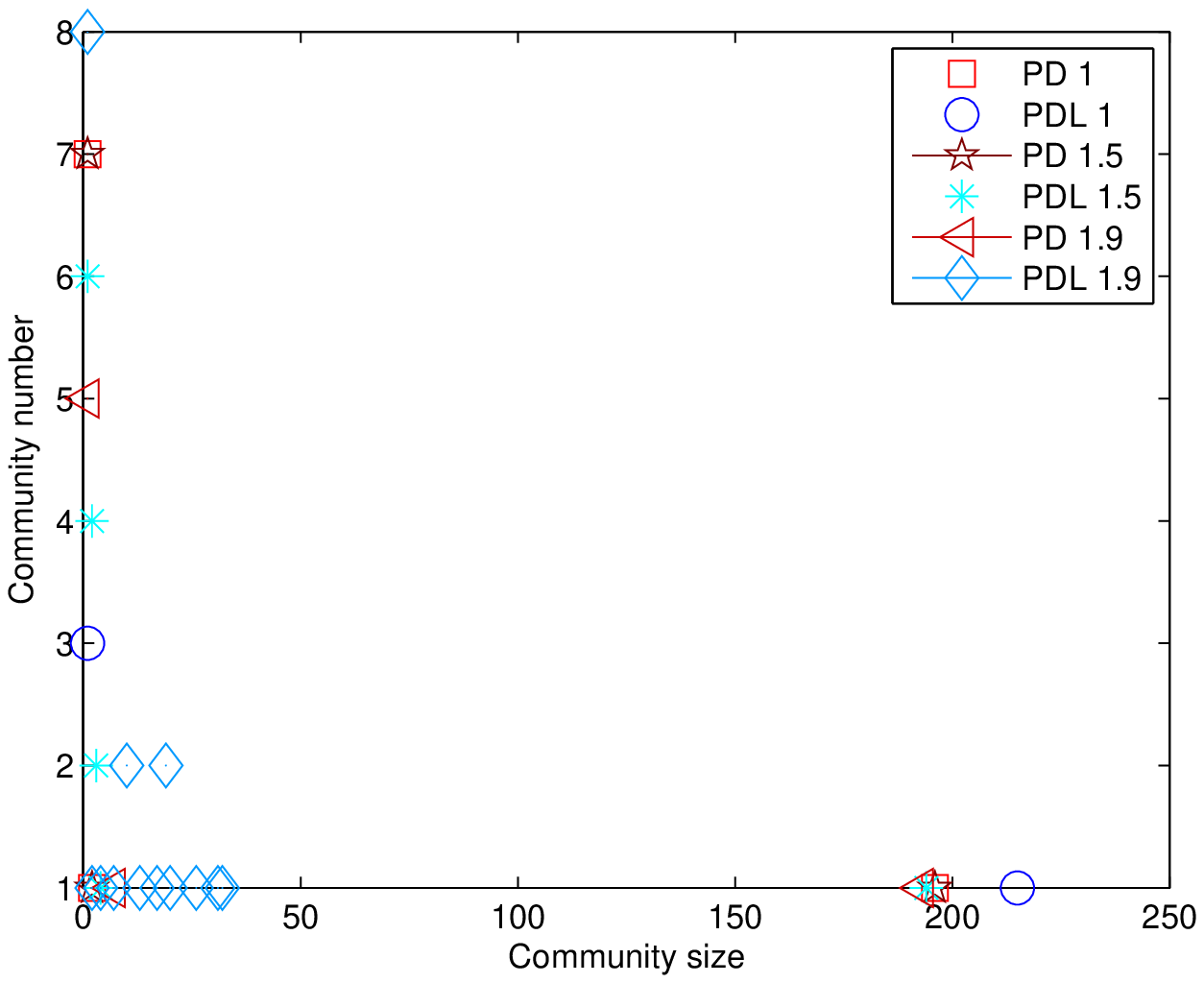}}
\subfigure[V Semester \label{}]
  {\includegraphics[width=0.5\textwidth]{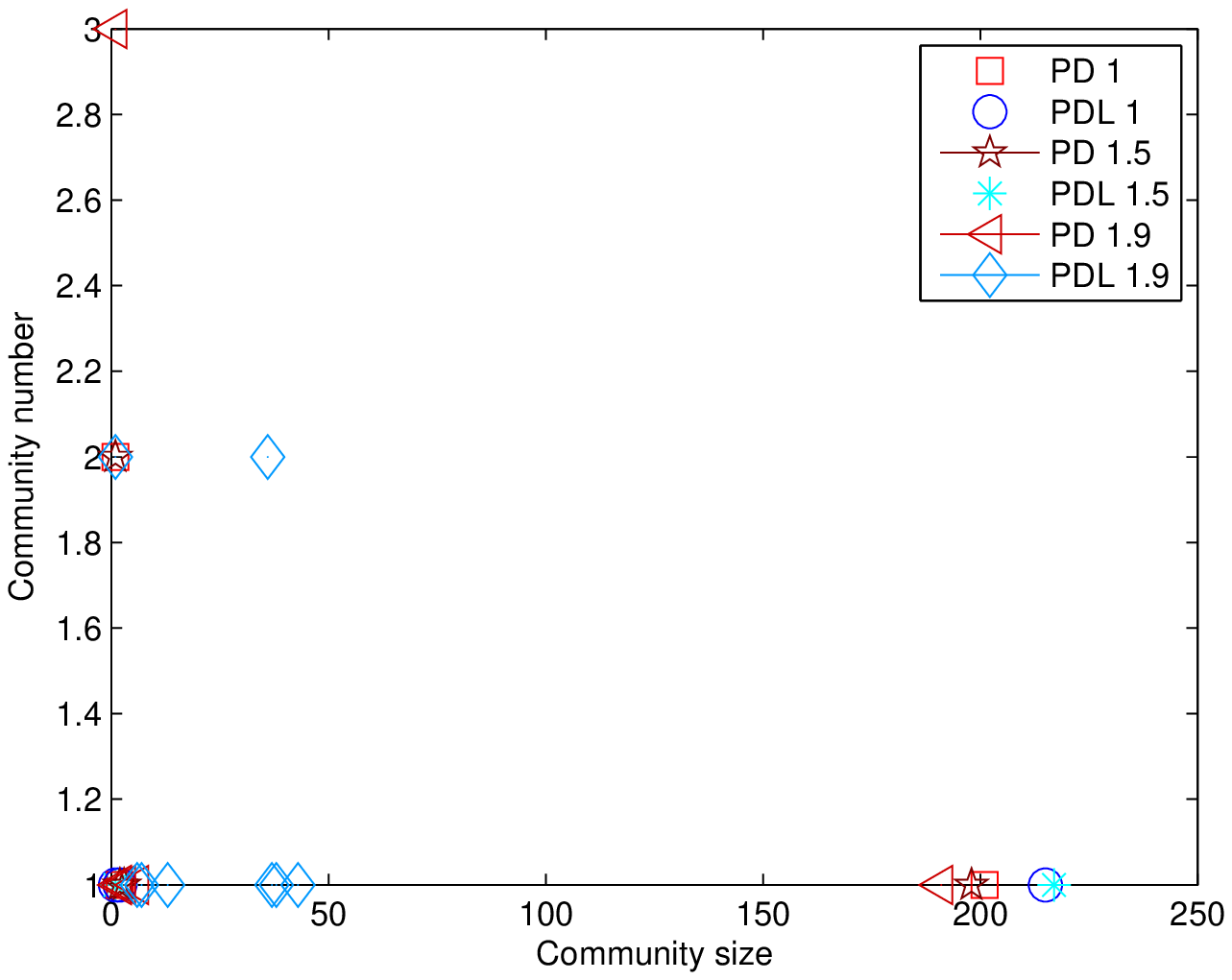}}
  \subfigure[VI Semester \label{}]
  {\includegraphics[width=0.5\textwidth]{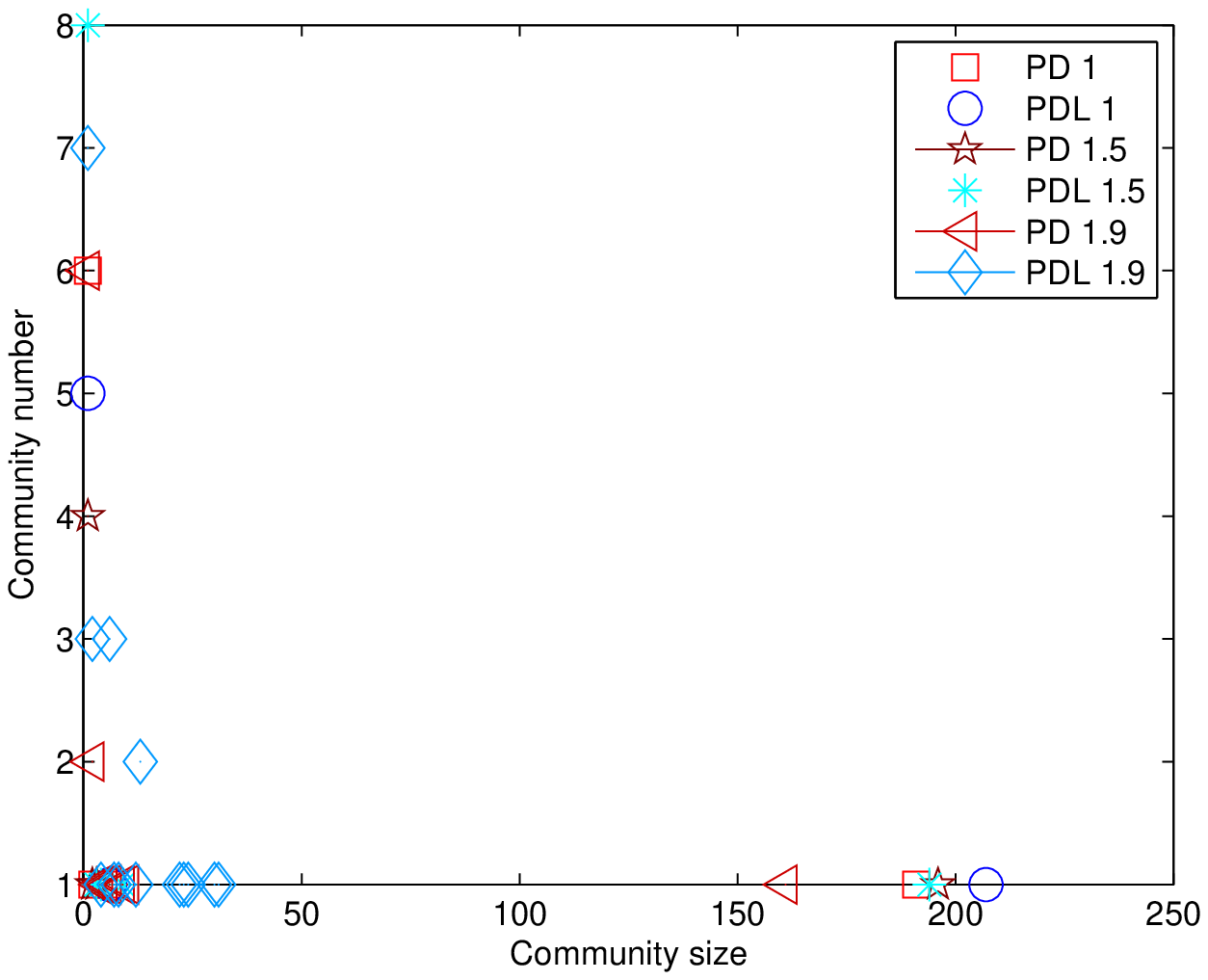}}

  \caption{Number of communities in which the two main parties PDL and PD are split and respective size for IV, V, and VI semesters, for different values of $\gamma$.}
\label{comm2}
\end{figure}

\section{Community structure}\label{modul}
In this section we apply network analysis techniques to the voting records of Italian Parliament to verify if the results obtained with the approaches employed in the previous sections are comparable when changing the analysis method. 
To this end we consider the binary matrix $B$ with $\delta=0.6$. This means that two Representatives are connected if they voted in the same way in at least  60\% of the overall roll calls.  The community structure of $\cal N$ can then be investigated by optimizing the well known concept of $modularity$ \index{modularity}\cite{NewmanGirvan2004}, based on the intuitive idea that a community should have more internal connections among its nodes than interconnections between its nodes and those in other communities.  Modularity is defined as  
\begin{equation}
Q=  \frac{1}{2r} \sum_{ij}(B_{ij} - \gamma \frac{k_i k_j}{2r})\delta(C_i, C_j) 
\end{equation}

where  $r$ is the number of edges in the network, $k_i$ is the degree of node $i$, $C_i$ is the community to which $i$ belongs, and $\delta(C_i, C_j)$ is 1 if nodes $i$ and $j$ belong to the same community, 0 otherwise. $\gamma$ is a resolution control parameter introduced by Granell et al. \cite{Granel2012} to overcome the resolution problem stated  in \cite{Fortunato2007a} and study community structure at multiple scales.  In fact it has been proved that the optimization of modularity has a topological resolution limit that depends on both the total size of the network and the interconnections of groups.  This implies that small,  tightly  connected clusters could not be found. Thus, searching for partitioning of maximum modularity, may lead to solutions in which important structures at small scales are not discovered. When $\gamma=1$ the equation reduces to the standard formulation of modularity \cite{NewmanGirvan2004}. 

We used an algorithm optimizing modularity  \cite{pizzuti2012} extended with the resolution parameter \index{resolution parameter}, and executed the method with three different values of $\gamma$: 1, 1.5, 1.9. The latter two values have been chosen to analyze the existence of sub-communities inside those obtained with $\gamma=1$ that cannot be found by optimizing modularity because of the resolution problem. \\
Figure \ref{mod} shows how modularity values vary during the seven semesters for all the three resolution parameters chosen. The figure clearly points out a sharp decrease of modularity in the 6th period and a drastic reduction in the 7th one. In order to better analyze the community structure detected by the algorithm,  Figures \ref{comm1} and  \ref{comm2} show the number of communities in which the two main parties PDL and PD have been split. We do not report the results for the other parties because their behavior is analogous to the coalition they belong. Since the size of the largest community is 218 (i.e. the number of PDL members), the first coordinate varies between 1 and 218. The second coordinate, for each value of $\gamma$, reports the number of subgroups of that size obtained by the algorithm. Figure \ref{comm1}(a) shows that,  with $\gamma=1$  PDL is grouped in a unique community, while PD is clustered in a big community of 190 members and other 14 members are split in 7 small communities. When  $\gamma=1.5$  the situation is almost the same. However, when $\gamma=1.9$, PD continues to have a big community of size 192, while PDL is split in 14 communities of size varying between 1 and 46. The very interesting result is that this behavior is maintained for all the semesters. Thus, while PD remains cohesive for all the semesters, independently of the $\gamma$ value, PDL  is divided into many subgroups since the first semester, when its degree of aggregation was considered very high, and as obtained with the other approaches described in the previous sections.  

Thus modularity allows a more deep analysis of the internal agreement of parties and can provide insights of early and unexpected changes a political party could encounter. 
Moreover, it affords an explicit and clear view of the steady fragmentation of the coalition endorsing the center-right government that culminated in its fall.

\begin{figure}[ht]
  \centering
  \subfigure[\label{}]
  {\includegraphics[width=0.4\textwidth]{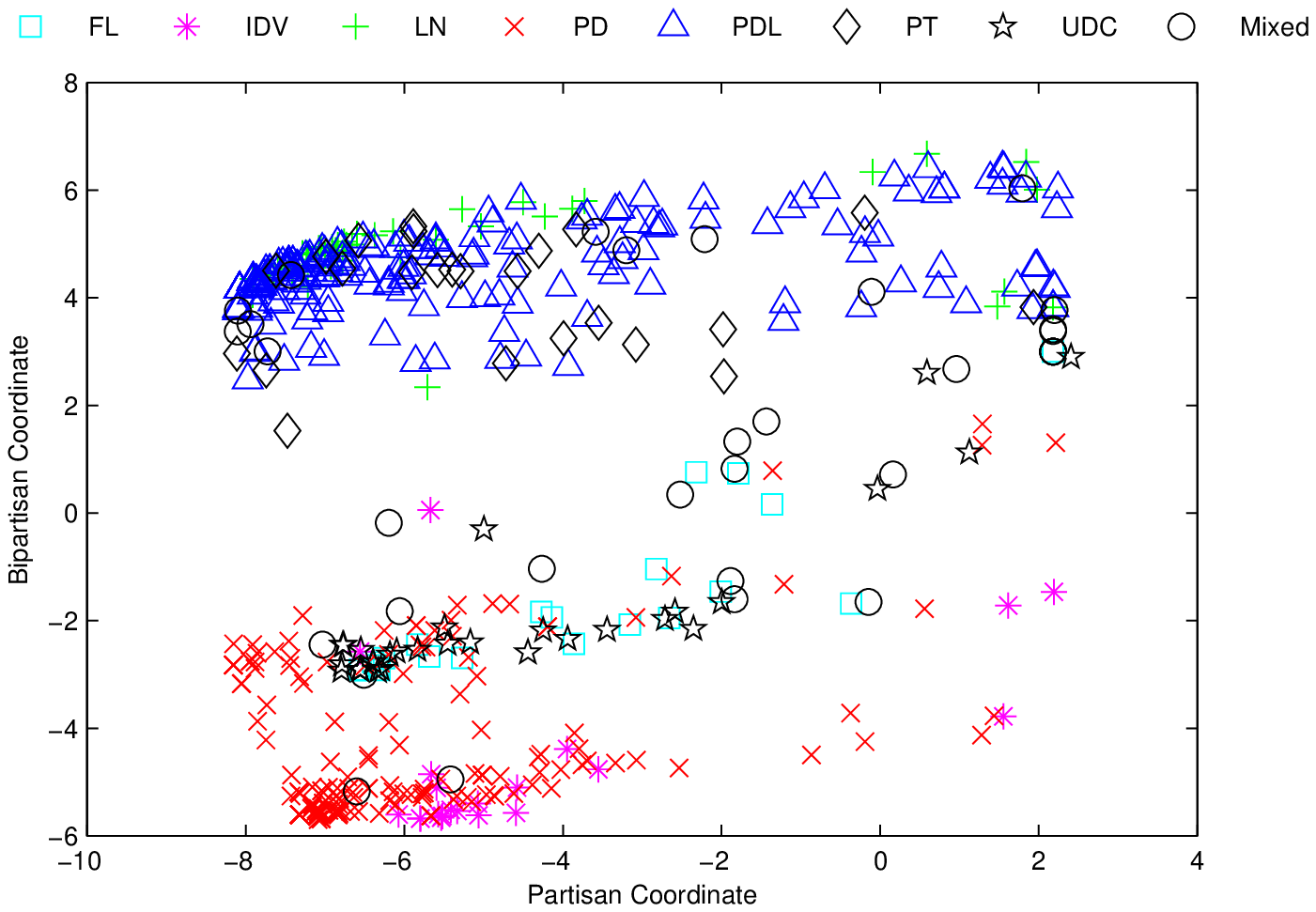}}
   \subfigure[\label{}]
  {\includegraphics[width=0.4\textwidth]{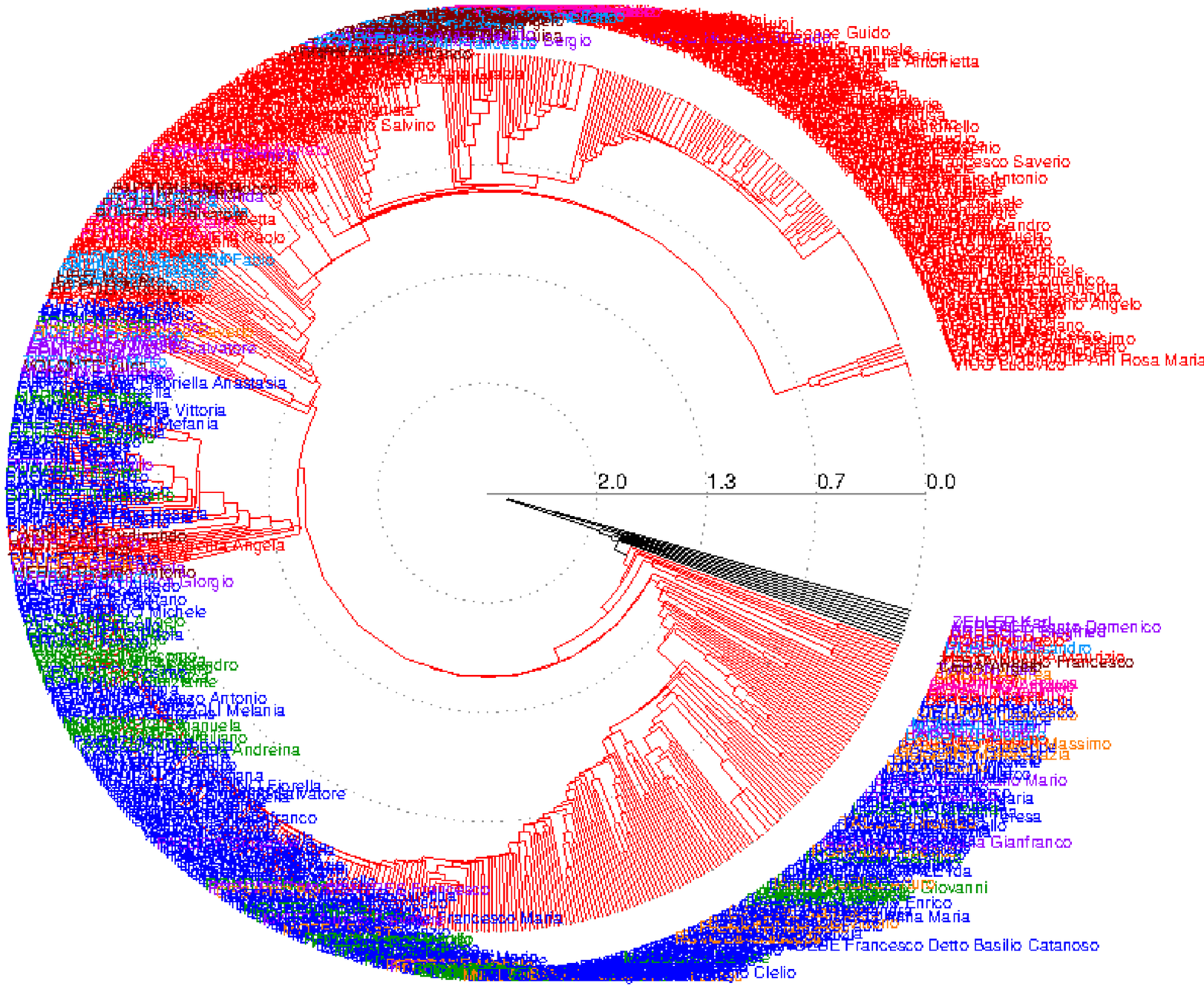}}
  \subfigure[\label{}]
  {\includegraphics[width=0.4\textwidth]{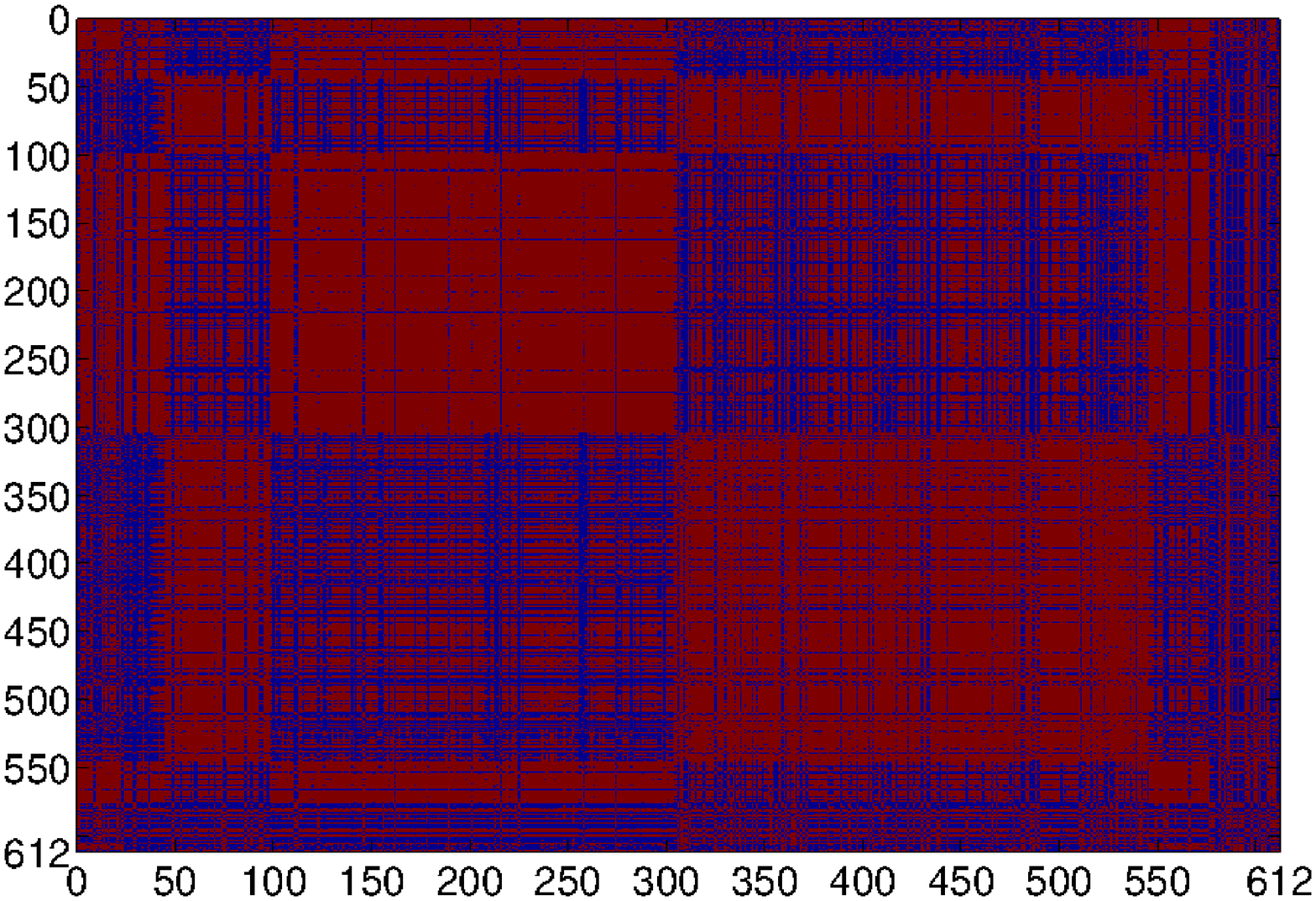}}
\subfigure[\label{}]
  {\includegraphics[width=0.4\textwidth]{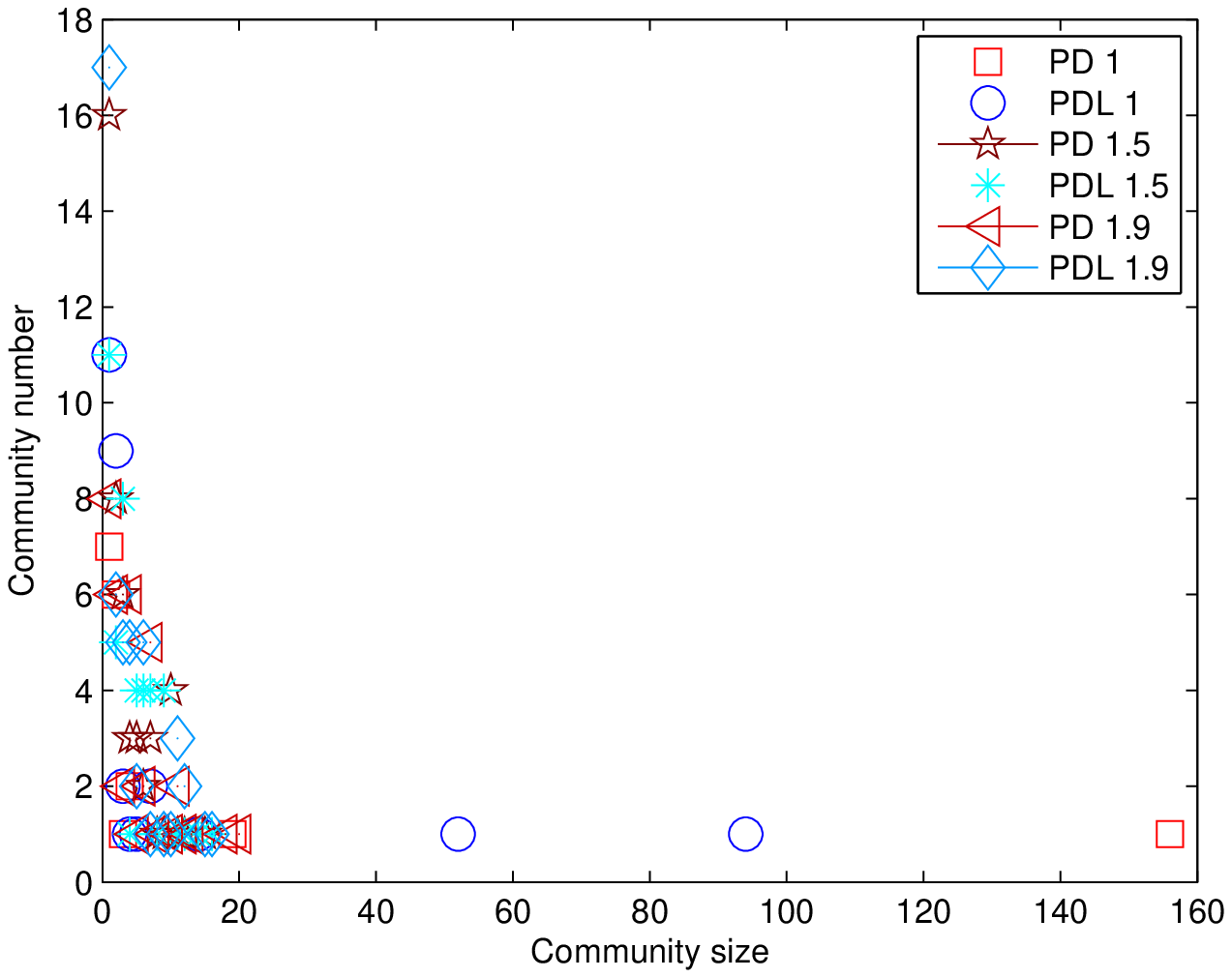}}

  \caption{Results obtained by applying \emph{SVD} (a), hierarchical clustering (b), visualization of the similarity matrix (c), and community detection (d) on the 7th semester.}
\label{lastsem}
\end{figure}

\section{The 7th Semester} \label{seven}
The analysis described in the previous sections mainly considered the first six semesters. We decided to separate the last semester because the voting behavior of Parliamentarians had an abrupt alteration, as testified also by the results obtained by all the employed methods. First of all, the number of voted measures is less than the fifth part of the other semesters. Furthermore, it happened that the political party organization completely disappeared, and each Parliamentarian voted independently of his group.  

Figure \ref{lastsem} gives a clear representation of this situation. In fact, the application of $SVD$
on this semester (Figure \ref{lastsem}(a)) shows a polarization of all the parties on the first coordinate, and distinguishes between center-left and center-right only on the bipartisan coordinate.  Hierarchical clustering returns a unique cluster including all the parties (Figure \ref{lastsem}(b)), and the visualization of the voting matrices (Figure \ref{lastsem}(c)) depicts high fragmentation. Finally, Figure \ref{lastsem}(d) shows that modularity optimization with $\gamma=1$ extracts a group of  156 and another of 19 members from PD, and two groups of 94 and 52 members from PDL. However these groups are clustered together, thus confirming the results of the other approaches. For higher values of $\gamma$, both parties are split in small groups of at most 20 Parliamentarians, and the communities found are constituted by members of almost all the political parties. 

It is worth to note that, as already pointed out, Figure \ref{mod} indicates an abrupt lowering of modularity value in the 7th semester \index{7th semester} that explains the loss of community structure.

\section{Related work}\label{rel}
The investigation of voting records with computational techniques is not new. One of the first paper is that of Jakulin and Buntine \cite{Jak2004}, where the authors analyzed the United States Senate in year 2003. They considered the Senate roll calls and the votes cast by each of the US Senators to compute a similarity matrix for every pairs of Senators, based on the Shannon's information theory concept of \emph{mutual information}  
\cite{Shannon1948}. The higher the mutual information between two Senators, the greater their similarity. Hierarchical clustering employed on their similarity matrix allowed to distinguish quite precisely between Republicans and Democrats. Furthermore, discrete blocks are identified, and similarity and dissimilarity among these blocks, with the aim of determining the voting influence of a single Senator, computed.   The authors observed that, though it is very difficult for a single Senator to influence final voting results because rarely a single vote changes the outcome of a roll call, once the blocks voting in a similar way are detected across a number of roll calls, the influence of changed behavior of a group can  be analyzed. In particular, two kinds of altered behavior have been considered: block abstention and block elimination. By using this approach, it was possible to obtain a list of roll calls for which it is deemed that the behavior of a block can affect the  outcome.

The same authors, with Pajala \cite{Pajala04}, analyzed the Finnish Parliament in year 2003. The Finnish Parliament is composed of 200 members elected for a four-year term. 
In 2003 elections changed the cabinet composition, thus Pajala et al. studied the cohesion of new political groups by computing the \emph{agreement index}  \cite{Hix2005}.
They found that the  groups composing the majority were more cohesive than the opposition groups. 
Moreover, they considered the roll calls and the votes cast by each of the Parliamentarians to compute a dissimilarity matrix between every pairs of Parliamentarians, based on Rajski's distance \cite{Rajiski1961}, that uses \emph{mutual information} and joint entropy. The lower the Rajski's distance between two Parliamentarians, the greater their similarity. 
They used the agglomerative hierarchical clustering algorithm \emph{Agnes} \cite{kauf1990} with the average linkage method, and built dendrograms. All the Parliamentarians were partitioned into clusters by the hierarchical clustering method. The results obtained showed that the analysis performed is able to capture the main characteristics of the Finnish Parliament. 
    
Another interesting study regarding the United States House of Representatives from 101st-108th Congresses has been done by Porter et al. \cite{Porter2007}. They defined bipartite collaboration networks from the assignments of Representatives to House committees and subcommittees. Each edge in the network between two (sub)committees has a weight which corresponds to the normalized interlock. The ��interlock�� between two committees is equal to the number of their common members. The normalization is obtained by considering the committee sizes, and dividing the interlock by the expected number of common members, if assignments were defined independently and uniformly at random. Then the hierarchical and modular structure of these networks, by using different community detection methods, has been investigated. Various methods of hierarchical clustering have also been executed. From the analysis, four hierarchical levels of clustering have been extracted: subcommittees, standing and select committees, groups of standing and select committees, and the entire House. The dendrograms revealed also an organization corresponding to groups of subcommittees inside larger standing committees. In order to perform an analysis of the obtained hierarchies in the House committee networks, authors used the modularity concept, modified to mine committee weighted networks. Instead of counting numbers of edges falling between particular groups, they counted the sums of the weights of those edges. This concept of modularity measures when a particular division of the network has more edge weight within groups than one would expect on the basis of chance, and it is used to evaluate the efficacy of the organizational grouping of the networks and to compare the dendrograms to each other. 
The community structure of the network of committees has been explored by using three other methods: two based on ��betweenness�� values computed on the full bipartite networks of Representatives and committees, and a local community detection algorithm for weighted networks. 
In this way, the authors identified connections between committees and correlations among committee assignments and Representatives' political positions. Changes in the network structure corresponded to change of Senate majority from Democrats to Republicans.
Finally, they applied $SVD$ to evaluate the House roll call votes. From this analysis, it was possible to observe as Democrats are grouped together, and are almost completely separated from Republicans.

Zhang et al. \cite{Zhang2008} studied the United States Congress by building bipartite networks for Members of Congress. In these ``bipartite" networks, there are two types of nodes: Congressman and bills, and a Member of Congress is linked by an edge to each sponsored or cosponsored bill.  By using information about the Congressional committee and subcommittee assignments, the authors created another kind of bipartite network where nodes are Representatives and committees/subcommittees, and an edge $(i,j)$ indicates the assignment of Representative $i$ to committee or subcommittee $j$. 
Each network is recursively partitioned in order to generate trees or �dendrograms� to assess its hierarchical structure. This process is able to discover communities of various sizes by iteratively clustering the legislators  by using the partitioning algorithm. Modularity evaluates the number of intra-community versus inter-community links for a given partition, consequently it has been adopted to quantify the growth in polarization in the U.S. Congress. In particular, during the considered period of 24 years, from the 96th to 108th Congresses, an increase in modularity has been obtained. This corresponded to an increase in party polarization of the Congress that caused the control by the Republicans of both chambers. Authors used also a multidimensional scaling technique called NOMINATE and singular value decomposition analysis. They showed that a matrix of roll call votes can be approximated by using two coordinates: a generic liberal-conservative dimension and a second �social� dimension. However, the same approaches demonstrated that multiple dimensions are needed to adequately approximate a matrix of cosponsorships. 
The adopted eigenvector methods detected large communities corresponding to known political cliques. It has been showed that Members of Congress with similar ideologies are clustered together in the identified communities.

Waugh et al.  \cite{WPFMP09}  evaluated  the polarization in the United States Congress by using also the concept of network modularity. Each node represents a legislator in the network and each edge is the level of agreement between two legislators in roll-call voting, indicating the average number of equal votes between them. Generally in a legislature, groups like parties contain strong connections between legislators within the same group but relatively weak connections between individuals in different groups.
Multiple community-detection algorithms have been employed on the similarity matrices of legislators to identify
groups that maximize the modularity inside each roll-call network for both the Senate and the House of Representatives. 
Modularity is adopted for measuring the degree of polarization, revealing the main political groups
and the divisions among them. A non-monotonic relationship between maximum modularity and a consequent majority party switch has been explored, demonstrating that the changes in majority are more likely when the modularity value is moderate, uncommon otherwise. In particular,  modularity values in Congress $t$ are used to predict modifications in the majority party for Congress $t+1$. A non-monotonic relationship between modularity and the stability of the majority party was found in both chambers of Congress. When modularity is low, a change in majority control seems to be less likely; at high levels of modularity, the minority cannot overcome majority cohesion. In both of these cases, it is infrequent to have majority-party switches. However, when modularity exhibits medium values, this corresponds to changes taking place for majority cohesion and to a less stable party system.This is called ``partial polarization" hypothesis. 

At the individual-level, some measures associated with modularity, called ``divisiveness" and ``solidarity" are computed to predict the reelection success for individual House members. The divisiveness measures the effect that each legislator could have on the aggregate polarization of his legislature by using roll-call adjacency matrices. About solidity, when its value is close to $1$, the legislator and community are strongly aligned. Performing this kind of analysis, authors found that divisiveness has a negative influence on reelection chances and that group solidarity has a positive influence. Furthermore, divisiveness is associated with decreased reelection probability, and the combination of divisiveness and solidarity has a significant positive impact on reelection. 

Macon et al. \cite{Macon2012} investigated the community structure of networks constructed from voting records of the United Nations General Assembly (UNGA). The UNGA was founded in 1946. Annual sessions from 1946 to 2008 have been considered and unanimous votes removed from the data, because they don't give information about the network structure of voting agreements and disagreements between countries. 
Three different networks have been defined. The first one is a weighted unsigned network of voting similarities, whose nodes are the countries and whose edges between pairs of countries are weighted by using an agreement measure. This represents the number of agreements on resolutions (yes-yes, no-no, or abstain-abstain) between the two involved countries.  The second kind of network is constructed by considering also the number of yes-no disagreements in the elements of the voting similarity matrix. The last kind of network is a signed bipartite network of countries voting for individual resolutions. 
By analyzing the resolutions with respect to the voting agreement, the authors were able to detect historical trends and changes in the United General Assembly community structure. In fact, observations appear to be consistent with the expected East-West split of the Cold War and the North-South division of recent sessions that has been detected by social scientists using qualitative methods.

\section{Conclusions}\label{concl}
The paper presented an investigation of the voting behavior of Italian Parliament in the last years by employing different computational tools. Though studies of this kind  exist for different political institutions from US and Europe, as far as we know, this is the first tentative  of exploring Italian Parliament with data mining and network analysis methods. 
We generated networks among the Parliamentarians at consecutive time periods and investigated  community structure at multiple scales.
By delving the voting records of Representatives, we were capable of characterizing the organizational structure of Parliament, and to discover latent information  contained. 
All the methods used showed to be effective at identifying political parties, and at providing insights on the temporal evolution of groups and their cohesiveness.
Future work aims at applying overlapping community detection methods to better uncover hidden collaborations among Parliamentarians of different political membership.

\end{document}